# High-Quality Ultra-Fast Total Scattering and Pair Distribution Function Data using an X-ray Free Electron Laser

Adam F. Sapnik,[a]* Philip A. Chater,[b] Dean S. Keeble,[b] John S. O. Evans,[c] Federica Bertolotti,[d] Antonietta Guagliardi,[e] Lise J. Støckler,[f] Elodie A. Harbourne,[g] Anders B. Borup,[f] Rebecca S. Silberg,[a] Adrien Descamps,[h] Clemens Prescher,[i] Benjamin D. Klee,[j] Axel Phelipeau,[k] Imran Ullah,[l] Kárel G. Medina,[l] Tobias A. Bird,[b] Viktoria Kaznelson,[f] William Lynn,[h] Andrew L. Goodwin,[g] Bo B. Iversen,[f] Celine Crepisson,[m] Emil S. Bozin,[n] Kirsten M. Ø. Jensen,[a] Emma E. McBride,[h] Reinhard B. Neder,[l] Ian Robinson,[o] Justin Wark,[m] Michal Andrzejewski,[p] Ulrike Boesenberg,[p] Erik Brambrink,[p] Carolina Camarda,[p] Valerio Cerantola,[p,q] Sebastian Goede,[p] Hauke Höppner,[r] Oliver S. Humphries,[p] Zuzana Konopkova,[p] Naresh Kujala,[p] Thomas Michelat,[p] Motoaki Nakatsutsumi,[p] Alexander Pelka,[r] Thomas R. Preston,[p] Lisa Randolph,[p] Michael Roeper,[k] Andreas Schmidt,[p] Cornelius Strohm,[k] Minxue Tang,[p] Peter Talkovski,[k] Ulf Zastrau,[p] Karen Appel,[p] and David A. Keen.[s]*

Email: afs@chem.ku.dk & david.keen@stfc.ac.uk.

---

## Abstract

High-quality total scattering data, a key tool for understanding atomic-scale structure in disordered materials, require stable instrumentation and access to high momentum transfers. This is now routine at dedicated synchrotron instrumentation using high-energy X-ray beams, but it is very challenging to measure a total scattering dataset in less than a few microseconds. This limits their effectiveness for capturing structural changes that occur at the much faster timescales of atomic motion. Current X-ray free-electron lasers (XFELs) provide femtosecond-pulsed X-ray beams with maximum energies of ~24 keV, giving the potential to measure total scattering and the attendant pair distribution functions (PDFs) on femtosecond timescales. Here, we show that this potential has been realised using the HED scientific instrument at the European XFEL and present normalised total scattering data for $0.35\text{Å}^{-1} < Q < 16.6\text{Å}^{-1}$ and their PDFs from a broad spectrum of materials, including crystalline, nanocrystalline and amorphous solids, liquids, and clusters in solution. We analyse the data using a variety of methods, including Rietveld refinement, small-box PDF refinement, joint reciprocal–real space refinement, cluster refinement, and Debye scattering analysis. The resolution function of the setup is also characterised. We conclusively show that high-quality data can be obtained from a single ~30 fs XFEL pulse for multiple different sample types. Our efforts not only significantly increase the existing maximum reported $Q$-range for an $S(Q)$ measured at an XFEL but also mean that XFELs are now a viable X-ray source for the broad community of people using reciprocal space total scattering and PDF methods in their research.

**a**: Department of Chemistry, University of Copenhagen, Universitetsparken 5, 2100 Copenhagen Ø, Denmark.
**b**: Diamond Light Source Ltd, Diamond House, Harwell Science & Innovation Campus, Didcot, Oxfordshire, UK.
**c**: Department of Chemistry, University Science Site, Durham University, South Road, Durham, DH1 3LE, UK.
**d**: Dipartimento di Scienza e Alta Tecnologia and To.Sca.Lab, University of Insubria, Como, Italy.
**e**: Istituto di Cristallografia and To.Sca.Lab, CNR, Como, Italy.
**f**: Center for Integrated Materials Research, Department of Chemistry and iNANO, Aarhus University, Langelandsgade 140, 8000 Aarhus C, Denmark.
**g**: Department of Chemistry, Inorganic Chemistry Laboratory, University of Oxford, South Parks Road, Oxford OX1 3QR, UK.
**h**: School of Mathematics and Physics, Queen's University Belfast, University Road, Belfast BT7 1NN, UK.
**i**: Institute of Earth and Environmental Sciences, University of Freiburg, Freiburg, Germany.
**j**: HUN-REN Wigner Research Centre for Physics, Konkoly-Thege Miklós út 29-33, 1121 Budapest, Hungary.
**k**: Deutsches Elektronen-Synchrotron DESY, Hamburg, Germany.
**l**: Friedrich-Alexander Universität Erlangen–Nürnberg, Staudtstrasse 3, D-91058 Erlangen, Germany.
**m**: Department of Physics, Clarendon Laboratory, University of Oxford, Parks Road, Oxford OX1 3PU, UK.
**n**: Center for Solid State Physics and New Materials, Institute of Physics Belgrade, University of Belgrade, Pregrevica 118, 11080 Belgrade, Serbia.
**o**: London Centre for Nanotechnology, University College London, London WC1E 6BT, UK.
**p**: European XFEL, Holzkoppel 4, 22869 Schenefeld, Germany.
**q**: University of Milano-Bicocca, Piazza della Scienza 4, 20126, Italy.
**r**: Helmholtz-Zentrum Dresden-Rossendorf (HZDR), Dresden, Germany.
**s**: ISIS Facility, Rutherford Appleton Laboratory, Harwell Campus, Didcot, Oxfordshire OX11 0QX, UK.

## Introduction

Atomic structure drives the properties that give materials their functionality. It is now widely recognised that understanding a material's properties often requires characterising both its average, long-range structure, and its local structure present over a much shorter range. This is especially clear in materials lacking long-range order, such as liquids and glasses, but is also the case in materials with ostensibly well-ordered periodic structures; deviations from the average structure locally can drive key physical behaviours (Keen & Goodwin, 2015; Simonov & Goodwin, 2020). Therefore, whether or not long-range order is present, the ability to study local atomic structure significantly benefits our understanding of a material's overall atomic architecture.

Total scattering, which encompasses both Bragg and diffuse scattering, along with its Fourier transform—the pair distribution function (PDF)—have been instrumental in the study of local atomic structure. Initially mainly applied to the study of liquids and glasses, these techniques were 'rediscovered' many years later for uncovering disorder within crystalline structures (Keen, 2020). Since then, total scattering and PDF analyses have been applied to a wide range of materials, primarily in equilibrium, including crystalline, nanocrystalline, amorphous, and liquid systems. The strength of these techniques lies in their universality, enabling the elucidation of atomic structure beyond the constraints of average structural analysis.

The pair distribution function (PDF) is a weighted probability of finding atom pairs at specific distances (Egami & Billinge, 2003). Key structural properties can be extracted directly from this function: peak positions give information about local bonding, peak areas provide information about coordination numbers, and peak widths reflect static and/or dynamic atomic disorder. The complexity of the atomic structure in modern functional materials is often reflected in the PDF as a series of overlapping peaks, making it increasingly difficult to interpret the data directly. As a result, model-based approaches are commonly used to derive meaningful structural insights from the PDF. Large-scale 'big-box' techniques sample ensembles of atoms which represent the material's structure (Tucker *et al.*, 2007), while 'small-box' techniques focus on interpreting the PDF within the framework of a crystallographic unit cell—an approach often referred to as 'real-space' Rietveld refinement (Farrow *et al.*, 2007). Regardless of the analysis method, obtaining high-quality total scattering data is essential for rigorous structural insights. A critical factor in data quality in total scattering experiments is measuring scattering to high momentum transfer ($Q$, where $Q = 4\pi \sin\theta/\lambda$) with sufficient signal-to-noise. The maximum $Q$ value ($Q_{max}$) in reciprocal space directly determines the resolution of the PDF in real space (Egami & Billinge, 2003). Additionally, during the generation of the PDF, the total scattering data are multiplied by $Q$ before performing the Fourier transform, which amplifies noise in the high-$Q$ region. Therefore, achieving high-$Q$ measurements with excellent signal-to-noise is essential for producing high-quality PDFs. Equally important is measuring the data in a manner that enables quantitative normalisation of the total scattering; stable detection systems, careful background measurements and calculations of contributions to the experimental data (Compton scattering, absorption, multiple scattering, *etc.*) must also be employed (Soper & Barney, 2011).

In recent years, there has been growing interest in studying the process of (dis)ordering itself, including phenomena such as amorphisation, crystallisation, nanoparticle formation and responses to ultrafast external stimuli (Terban & Billinge, 2022; Lindahl Christiansen *et al.*, 2020; Proffen, 2006). These investigations demand not only the high real-space resolution determined by $Q_{max}$ but also the ability to capture temporal changes, with timescales ranging from several hours to femtoseconds. Some of the fastest changes happen in key technologies, from quantum materials and ultrafast memory to photovoltaics and spintronics (Yasuda *et al.*, 2024; Jiang *et al.*, 2024; Afanasiev & Kimel, 2023; Guo *et al.*, 2024). Many transformative phenomena in materials science—such as photoinduced phase transitions, non-thermal melting, resistive switching, and ultrafast demagnetisation—unfold on femtosecond to picosecond timescales (Schmid *et al.*, 2024; Rousse *et al.*, 2001; Bigot *et al.*, 2009; Pan *et al.*, 2022; Weißenhofer & Oppeneer, 2024). Capturing these events in real time requires tools capable of probing atomic positions and electronic configurations with temporal resolution that matches the time over which these nonequilibrium states exist. These 'fleeting moments' might govern key functional behaviour, such as light-matter interactions in perovskites or reveal switching mechanisms in phase-change memory materials, or are the only time that highly transient phases exist during shock-wave compression experiments.

Synchrotron-based diffractometers, with their high X-ray energy and brilliance, are the instruments of choice for X-ray total scattering measurements. They routinely achieve $Q_{max}$ values of around 25 $Å^{-1}$, enabling high real-space resolution. However, synchrotron measurements typically take a few minutes. Recent advancements have further reduced measurement durations, with high-quality data obtainable in seconds. In some cases, sub-second data acquisition has been achieved, with time resolutions as short as 3 ms (Magnard *et al.*, 2023), however, such measurements are not common, and typical measurement times are still on the order of seconds to minutes. Currently, although synchrotrons are the best source for X-ray total scattering, offering an optimal balance between exceptional real space resolution and good temporal resolution, they cannot provide the ultra-fast capabilities argued for in the previous paragraph.

X-ray free-electron lasers (XFELs) are the pinnacle of modern X-ray sources. Building on concepts developed in the 1970s, XFELs use self-amplified spontaneous emission within a long undulator to generate an exceptionally brilliant and coherent X-ray laser beam from high-energy electrons (Georgescu, 2020). These facilities offer unparalleled brilliance and temporal resolution. The first XFEL, the Linac Coherent Light Source (LCLS) in the United States, became operational in 2009 (Emma *et al.*, 2010). Since then, five additional XFEL facilities have been established in Japan (Ishikawa *et al.*, 2012), South Korea (Kang *et al.*, 2017), Germany (Decking *et al.*, 2020), Switzerland (Prat *et al.*, 2020), and China (Liu, 2022). The European XFEL (EuXFEL) began operations in Hamburg, Germany, in 2017 (Tschentscher *et al.*, 2017). It accelerates electrons over a 1.7 km path, achieving energies of over 17.5 GeV. The facility also boasts a high repetition rate of 27000 pulses per second, with individual pulse durations shorter than 100 fs.

XFELs have opened new frontiers in physics, particularly matter under extreme conditions and materials chemistry, by enabling 'molecular movies' that capture atomic movements with extraordinary temporal precision. Current XFEL setups have excelled in femtosecond serial crystallography and pump–probe experiments, primarily focusing on Bragg diffraction, alongside coherent diffraction imaging and spectroscopy (Støckler, Krause *et al.*, 2023; Nakano *et al.*, 2017; Obara *et al.*, 2017). Total scattering measurements with PDF analysis were first demonstrated at XFELs in the 2000s. One early pump-probe study of gold trimer formation in solution achieved a $Q_{\max}$ of ~6.5 Å$^{-1}$ for difference structure factors $\Delta S(Q)$ (*i.e.* $S(Q)$ post-pumping minus $S(Q)$ pre-pumping) using an X-ray energy of 15 keV (Kim *et al.*, 2015). While this was a technological milestone, the spatial resolution of such data was comparable to that achieved with X-ray diffractometers from 100 years ago. Since then, progress to improve the real-space resolution has been limited with $Q_{\max}$ values of 7 Å$^{-1}$ and, at best, ~9 Å$^{-1}$ reported for total scattering structure factors, $S(Q)$, for shock compression measurements (see recent studies using 18 ke X-rays of liquid carbon (Kraus et al., 2025) and liquid tin (Gorman *et al.*, 2024)), respectively, or $Q_{\max}$ ~8 Å$^{-1}$ from liquid-jet nanoparticle suspensions in (Støckler, Christensen *et al.*, 2023). This results from low X-ray energies compounded by difficulties in normalising data from taxing experimental setups with limited $Q_{\max}$. To the best of our knowledge, the current $Q_{\max}$ record for the more straightforward difference structure factor, $\Delta S(Q)$, at an XFEL is 12.6 Å$^{-1}$ (corresponding to a real-space resolution of ~0.25 Å) (Griffiths *et al.*, 2024). While this resolution is adequate for simple structural transitions, it remains insufficient for many advanced studies and falls short of what Mo or Ag source diffractometers can achieve. To fully leverage the temporal capabilities of XFELs for total scattering applications, it is essential to improve the real-space resolution of any potential PDF instrumentation. Achieving this goal involves addressing three primary challenges: arranging detectors to mitigate the lower-than-ideal X-ray energies of XFELs, improving the experimental environment to reduce backgrounds and designing the experimental protocols to maximise the efficiency of the measurements.

This work presents our efforts to achieve high-quality, ultra-fast total scattering and PDF data at the High Energy Density (HED) scientific instrument at the European XFEL. Here, we describe the experimental setup in detail, including the use of a tilted detector geometry designed to maximise counting statistics in the high-$Q$ region. Additionally, we outline the data processing and reduction pipeline, ensuring precision and efficiency. To demonstrate the versatility of this setup, we present measurements across a diverse range of materials applications, highlighting its broad utility and potential for using these methods at XFELs for groundbreaking insights.

# Methods

## Experimental Setup

Our experiments used the High Energy Density (HED) scientific instrument at the European XFEL in Hamburg. The experiments were carried out in interaction chamber one (IC1), which operates under a $< 10^{-4}$ mbar vacuum, thus eliminating background contributions from air scattering. General details about IC1 and HED are given in (Zastrau *et al.*, 2021). We used a Varex 4343CT detector (housed inside a vacuum-tight air pocket with a 400 μm thick Al window in front of the active face of the detector) and a Jungfrau detector (Zastrau *et al.*, 2021), placed to maximise the $2\theta$ coverage and providing a $Q$ range of ~0.35 to 16.6 $\text{Å}^{-1}$ with an incident X-ray energy of 24.075 keV and beam energy of ~150 μJ (**Figure 1a**). The maximum $Q$ results in an approximate real-space resolution of 0.19 Å. The energy bandwidth was estimated at around 50 eV. Beryllium compound refractive lenses installed 114 m upstream of the target chamber centre (or sample position) of IC1 provided a focused 17 μm (vertical) by 55 μm (horizontal) X-ray beam of low divergence on the sample. The beam size was determined using round edge scans; the vertical dimension is likely to be close to the true beam size of a single pulse, whereas the larger horizontal value reflects the greater beam jitter in this direction and/or an imperfectly adjusted beam bender component. The beam size was selected to be compatible with other instrumentation in the setup; specifically, it was chosen to be smaller than the optical pump laser beams envisaged for future experiments. We also added a set of four tungsten-blade clean-up slits (200 μm × 200 μm) and a stainless steel plate with pinhole upstream of the sample holder to reduce background scattering (**Supplementary Figure 1**).

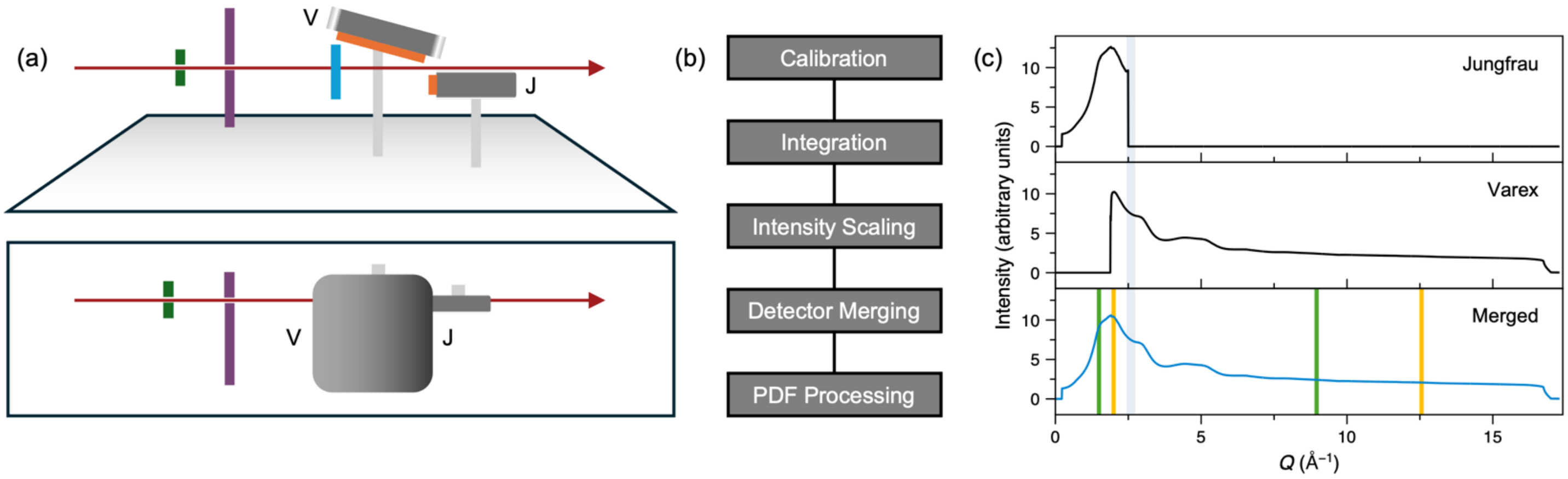

**Figure 1** (a) Schematic of the main diffractometer components (upper: side view, lower: plan view). The X-rays emerge from the standard HED IC1 components and pass into the IC1 vacuum chamber. They then pass through the clean-up slits (green), steel shield (purple), sample in a vertical holder (blue) and below the Varex (labelled 'V') and above the Jungfrau (labelled 'J') detectors (active areas in orange) before exiting the IC1 chamber. The Varex detector is not centred above the X-ray beam because of space constraints. The technical drawing of the setup can be found in **Supplementary Figure 1**. (b) Outline of the data reduction pipeline. (c) Example of processed data from the two detectors and the fully merged data, using water in a fused silica capillary as an example. (The normalised $S(Q)$ for water from these data is shown in Figure 10.) Recent $Q_{\min}$ and $Q_{\max}$ values for XFEL measurements of $S(Q)$ and $\Delta S(Q)$ are shown as green (Gorman *et al.*, 2024) and yellow (Griffiths *et al.*, 2024) vertical lines, respectively. The region over which merging is performed is shaded in grey; data at $Q$-values above and below this shaded region were not used for the Jungfrau and Varex detectors, respectively.

The Varex 4343CT detector has a 2880 × 2880 array of 150 μm square pixels, and is capable of operating at 10 Hz, the pulse train frequency of the European XFEL. It was placed above the straight-through beam at an angle of ~15° to the horizontal, covering ~9° < 2$\theta$ < ~89° in the vertical plane. With this tilted detector arrangement, which follows the work described in (Burns *et al.*, 2023), the position of nearest incidence of the detector is 171 mm from the sample at a 2$\theta$ of ~75°. The main advantages of this detector arrangement are threefold. Firstly, the detector is closest to the sample at higher scattering angles such that the detector solid angle is maximised where the sample scattering is weakest. Secondly, the detector is further from the sample at low-2$\theta$, increasing the detector's resolution for the low-angle Bragg peaks and reducing the likelihood of detector saturation. Thirdly, the range of scattering angles covered is significantly larger than if the detector was placed in the standard manner perpendicular to the X-ray beam. On account of the Varex detector's 16-bit ADC readout, for certain strongly scattering samples, we needed upstream attenuation of the primary beam in order to avoid saturation.

The Jungfrau detector is a smaller two-dimensional detector with a 1024 × 512 array of 75 µm square pixels. It has an automatic gain switching feature for each pixel, which allows it to detect single photon events up to high signal levels by gain switching to cover a broad range of count rates (Redford *et al.*, 2018, 2020). The detector is also in an air pocket to operate in vacuum environments (Zastrau *et al.*, 2021). It was placed below the straight-through beam 407 mm behind the sample and at an angle of ~90° to the horizontal, covering ~1° < 2$\theta$ < ~14.5° in the vertical plane. This detector was added to provide data in the low-$Q$ region since the minimum $Q$ possible with the Varex detector inside its vacuum pocket was ~2 Å$^{-1}$.

The HED sample scanner accommodates EUCALL (Appleby *et al.*, 2017; Prencipe *et al.*, 2017) standard sample holders, which consist of an outer and inner frame. The inner frame is compatible with variable target mount plates, which are required for flat sample plates or capillaries, the latter of lengths ~40 mm and ~70 mm. Capillaries were glued or taped directly to the sub-frames, enabling various diameter capillaries to be used and allowing for flexible sample mounting options. The sample scanner was mounted vertically (**Supplementary Figure 2**), and the accessible area on the sample plate was 100 mm × 100 mm, even with the top edge of the Varex detector air pocket sitting above the sample scanner. The focal point of a camera coincident with the beam was used to place the samples at the target centre of IC1.

## Data Processing

### Overview

We developed a robust data reduction pipeline to achieve high-quality data suitable for PDF analysis. This process involved the following steps (as represented in **Figure 1b**) and explained in more detail in the sections below. We applied this process to both the mean of selected pulse trains within a run and to those pulse trains individually. Note that for the majority of the experiments, each train only contained a single pulse, which matches the maximum readout time of the Varex detector.

1. Azimuthal integration of the data from the individual detectors to yield the intensity as a function of $Q$, $I_x(Q)$
2. Scaling and merging of the individual $I_x(Q)$ to form a merged $I_{merged}(Q)$
3. Subtraction of background and Fourier transform to calculate the pair distribution function

**Calibration**

Detector calibration was performed using pyFAI, specifically using the *pyfai-calib2* tool (Ashiotis *et al.*, 2015). A thin layer of NIST $CeO_2$ SRM 674b powder sandwiched between two Kapton films was used as the calibrant for both detectors to reduce sample thickness effects. The primary beam energy in the calibration fitting was set to 24.075 keV (a wavelength of 0.51499 Å), and both the energy and the inclination angle of the detector around the incident beam direction were not allowed to refine – the latter to ensure the correct polarisation correction was applied.

**Integration**

The azimuthal integration of the two detectors was performed using pyFAI to convert from a matched pair of images to a pair of $I_x(Q)$. For both detectors, we determined our own bad pixel masks using a combination of algorithmic and manual processes and applied a correction to account for the effective sensor thickness. In both cases, we also normalised each pixel to absolute solid angle before integration to simplify the combination of the signals from the two detectors. Additionally, we normalised each dataset to the measured intensity on a diagnostic silicon diode upstream of the sample to account for pulse-to-pulse variation in intensity. In cases where primary beam attenuation above 90% was utilised (see above), the diode signal was too weak to be used as a reliable intensity normalisation, and so no normalisation was used.

The Varex detector exhibits some features which require additional corrections. After subtraction of the signal arising from the dark current in the detector (the 'dark image'), we also applied intensity corrections to account for the attenuation caused by the aluminium window and the detector response, *i.e.* a flat field correction. Data from the Jungfrau detector were also corrected for the aluminium window. Details of how these corrections were performed can be found in the **Supplementary Information & Supplementary Figure 3**.

**Merging**

To combine the Varex $I_v(Q)$ and Jungfrau $I_j(Q)$ data we first scale $I_j(Q)$ to minimise the difference between it and $I_v(Q)$ in the region where the two overlap (between 1.89 and 2.49 Å$^{-1}$), applying both an offset and a scale factor *via* a least squares minimisation (**Figure 1c**). This step is required because of the different sensor efficiencies of the two detectors. Once on the same scale, a smooth Q-dependent weighting scheme was applied using an error-function-based ramp. The error function, centred at $Q = 2.2$ Å$^{-1}$ and with a width of 0.1 Å$^{-1}$, scales the overlapping regions of the two detectors. Overlapping intensities were merged, and normalisation/count statistics were recomputed accordingly.

**Monitoring of the dark image validity**

In the knowledge that the signal arising from the dark current in the Varex (the 'dark image') is time varying and very sensitive to temperature, in each sample run, we pre-checked the applicability of the stored dark image. We did this by collecting a number of frames without accepting X-rays into the interaction chamber and asserting that, after subtraction of the current dark image, the mean intensity should be zero and that the standard deviation of the intensities should be below a threshold of 15 (**Figure 2**).

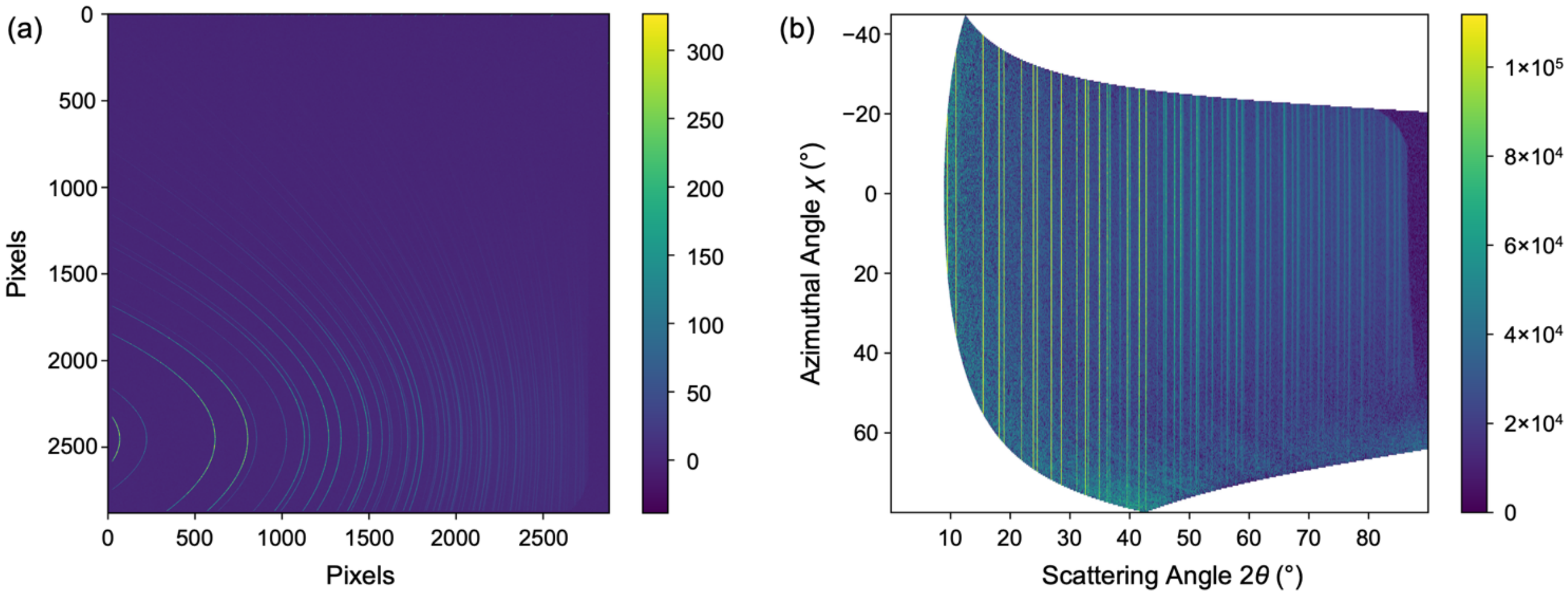

**Figure 2** Varex detector images from $CeO_2$ powdered sample. **(a)** Detector image and **(b)** data plotted as a function of scattering and azimuthal angles, showing vertical lines of intensity corresponding to Bragg reflections. See **Supplementary Figure 4** for the equivalent data from the low-angle Jungfrau detector.

## Correction for sample offset

A sample offset correction was applied by deriving the general expression for a flat-plate detector inclined at an angle α to the incident beam (see **Eq. 1**) for small offsets. Note that in the case of $\alpha = 90°$, the correction simplifies to the expression derived by Hulbert *et al.* (Hulbert & Kriven, 2023). $\delta$ is the $2\theta$ offset, $S$ is sample displacement along the beam, $R$ is sample-to-detector distance, and $\alpha$ (here ~15°) is the inclination angle of the detector to the incident beam.

Equation 1:

$$\delta = \frac{S\,\sin(2\theta)\,\sin(2\theta+\alpha)}{(R-S)\sin(\alpha)+S\cos(2\theta)\sin(2\theta+\alpha)}$$

## Calculating the total scattering structure factor and pair distribution function

Following the data processing steps, PDFs were generated from the total scattering *via* normalisation and Fourier transformation using either GudrunX (Soper, 2011) or PDFgetX3 (Juhás *et al.*, 2013), following standard procedures. In GudrunX, the raw total scattering data were corrected for background, multiple scattering, container scattering and Compton scattering, and for fluorescence and absorption effects (Soper, 2011). In PDFgetX3, where the data are corrected using an *ad-hoc* approach, the following parameters were used: $Q_{\mathrm{min}}$: 0.9–2.2 $Å^{-1}$, $Q_{\mathrm{max}}$: 15–16.6 $Å^{-1}$, $Q_{\mathrm{max,inst}}$: 16–16.6 $Å^{-1}$, $r_{\mathrm{poly}}$: 0.9–1.86 (Juhás *et al.*, 2013).

## Sample preparation

Si, $CeO_2$ and $LaB_6$ were obtained as certified standard reference materials (SRMs) from the US National Institute of Standards and Technology (NIST). Titanium dioxide P25, silver behenate, and ammonium metatungstate were purchased from Sigma Aldrich. The metallic glass ($Fe_{78}B_{13}Si_9$) was purchased from Goodfellow as a 25 µm-thick foil. Silica glass was measured in the form of a fused silica capillary, coated in a thin polyimide layer, with inner and outer diameters of 700 and 850 µm, respectively. All chemicals were used without further purification. Flat plate samples were prepared by dispersing powders onto Kapton tape. Capillary samples were measured in fused silica (0.7 mm inner diameter) coated with a thin polyimide layer or quartz (0.5 mm inner diameter) capillaries.

## Results and Discussion

We applied our methodology to a diverse set of samples to show that quantitative data over a wide range of $Q$ can be obtained across different material types (**Figure 3**). We investigated highly crystalline powders (Si, $CeO_2$, $LaB_6$ and Ag behenate), a nanocrystalline powder ($TiO_2$), amorphous solids (silica glass and an $Fe_{78}B_{13}Si_9$ metallic glass), a pure liquid (water) and an atomic cluster in solution (aqueous tungsten Keggin, $[W_{12}O_{40}]^{6-}$).

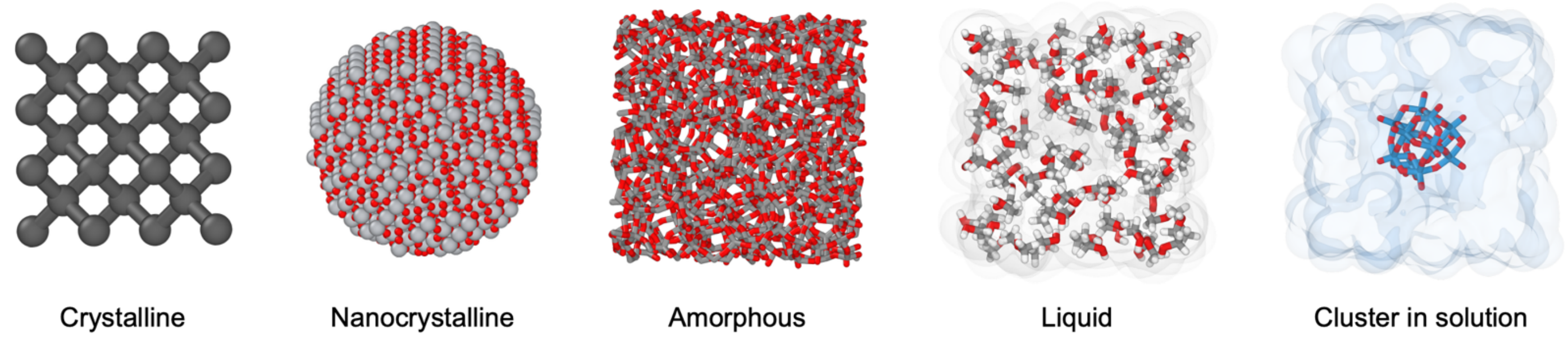

**Figure 3** Overview of the material types measured in this study.

To determine the diffractometer characteristics, we measured highly crystalline powders using capillary and flat-plate geometries. These were first used to determine the positions of the detectors relative to the sample, and then, because they exhibit minimal sample-induced peak broadening, the instrumental resolution function (IRF).

*Crystalline standards: NIST Si 640b and $CeO_2$ 674*

A quartz capillary loaded with NIST Si 640b SRM was measured using a 1-pulse per train acquisition, averaged over ~1700 pulses. The resulting diffraction pattern exhibited narrow, symmetric peaks with low background and an excellent signal-to-noise ratio, despite the inclined detector geometry, which places the detector close to the sample and distorts the Debye-Scherrer powder rings. The data spanned a $2\theta$ range of 1.1 to 86.0° (0.23 to 16.6 Å$^{-1}$). Even at the highest angles, Bragg peaks remained sharp with good signal-to-noise (the latter due to the increasing solid angle with increasing $2\theta$).

Rietveld refinement using TOPAS academic (Coelho, 2018) yielded a high-quality fit ($R_p$ = 6.392%, $R_{Bragg}$ = 3.045% ) and an atomic displacement parameter, $B_{iso}$(Si), of 0.328(7) Å$^2$ using a fixed lattice parameter (certified by NIST) of 5.43094 Å (**Supplementary Figure 5a**). The peak shapes were well-described using a standard Pseudo-Voigt profile, and background modelling was performed using a 13$^{th}$-order Chebyshev polynomial. The sample offset was determined to be −0.039(1) mm relative to the nominal diffractometer centre (see **Eq. 1** in **Methods**).

Following normalisation, the NIST Si 640b reciprocal space data were Fourier transformed into a PDF using PDFgetX3 (Juhás *et al.*, 2013) using a $Q$-range of 1.7 to 16 Å$^{-1}$. The resulting PDF displayed sharp peaks with minimal damping at high-$r$, confirming negligible instrumental contributions. Small-box Rietveld refinement in PDFgui (Farrow *et al.*, 2007) produced a good fit ($R_p$ = 13.7%), a lattice parameter of 5.4316(7) Å, and a $B_{iso}$(Si) of 0.5(1) Å$^2$ (**Supplementary Figure 5b**). The instrumental parameters, $Q_{damp}$ and $Q_{broad}$, were determined to be 0.006(4) and 0.009(4), respectively. The difference curve had some oscillatory behaviour, possibly an artefact from the Fourier transform of the merged region of the data (which occurs in the region between 1.89 and 2.49 $Å^{-1}$). A joint refinement of powder diffraction and PDF data in TOPAS Academic (Coelho, 2018) yielded similar results ($R_w$ = 6.72%), with a $B_{iso}$(Si) of 0.42 $Å^2$—close to the average of the values obtained from refinements using TOPAS academic of the powder diffraction pattern and PDF data separately (**Figure 4**).

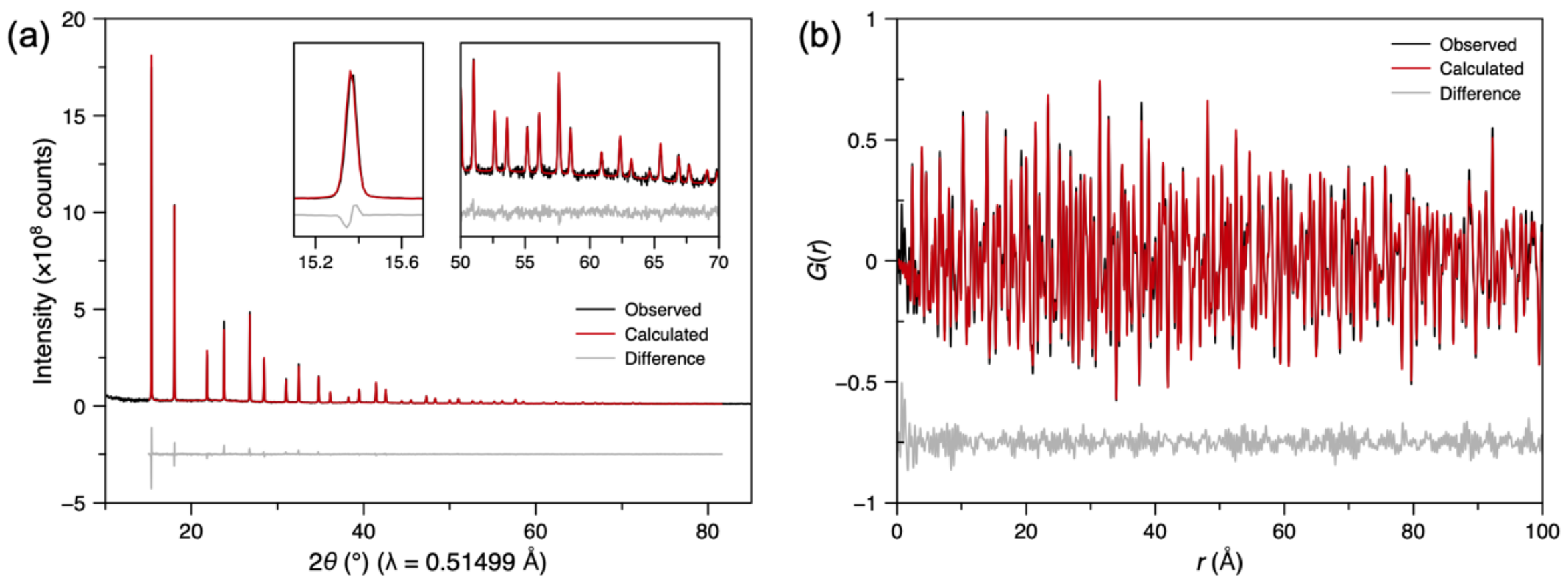

**Figure 4 (a)** Rietveld refinement of averaged normalised powder diffraction data of NIST Si 640b SRM loaded in a quartz capillary. Insets show enlarged regions of the data at low- and high-2*θ*. **(b)** Small-box 'real-space Rietveld' refinement against the PDF obtained from the data in (a). The Rietveld and small-box PDF refinements shown in (a) and (b), respectively, were carried out simultaneously using TOPAS academic software (Coelho, 2018).

To assess the feasibility of ultra-fast total scattering measurements, NIST $CeO_2$ 674 SRM powder in a fused silica capillary was measured using a single XFEL pulse (~30 fs) with only 10% of the full X-ray beam intensity. Remarkably, despite the extremely short acquisition time, around 16 orders of magnitude faster than a typical synchrotron measurement, the diffraction pattern exhibited sharp Bragg peaks and a high signal-to-noise ratio (**Figure 5a**). Rietveld refinement of the single-pulse data yielded a mean $R_p$ of 8.26(1.14)%, $R_{wp}$ of 10.85(1.56)%, a lattice parameter of 5.4079(0.0018) Å, and $B_{iso}$ (Ce) of 0.25(0.04) $Å^2$ and $B_{iso}$ (O) of 0.51(0.17) $Å^2$, using a Thomson-Cox-Hastings pseudo-voigt (TCHZ) peak profile. Here we give the values' mean and standard deviation (in brackets) obtained from refinements of 74 single pulse measurements. These values compare reasonably well to previously reported atomic displacement parameters of 0.292 and 0.395 $Å^2$ for $B_{iso}$ (Ce) and $B_{iso}$ (O), respectively, obtained from synchrotron data (Yashima & Takizawa, 2010).

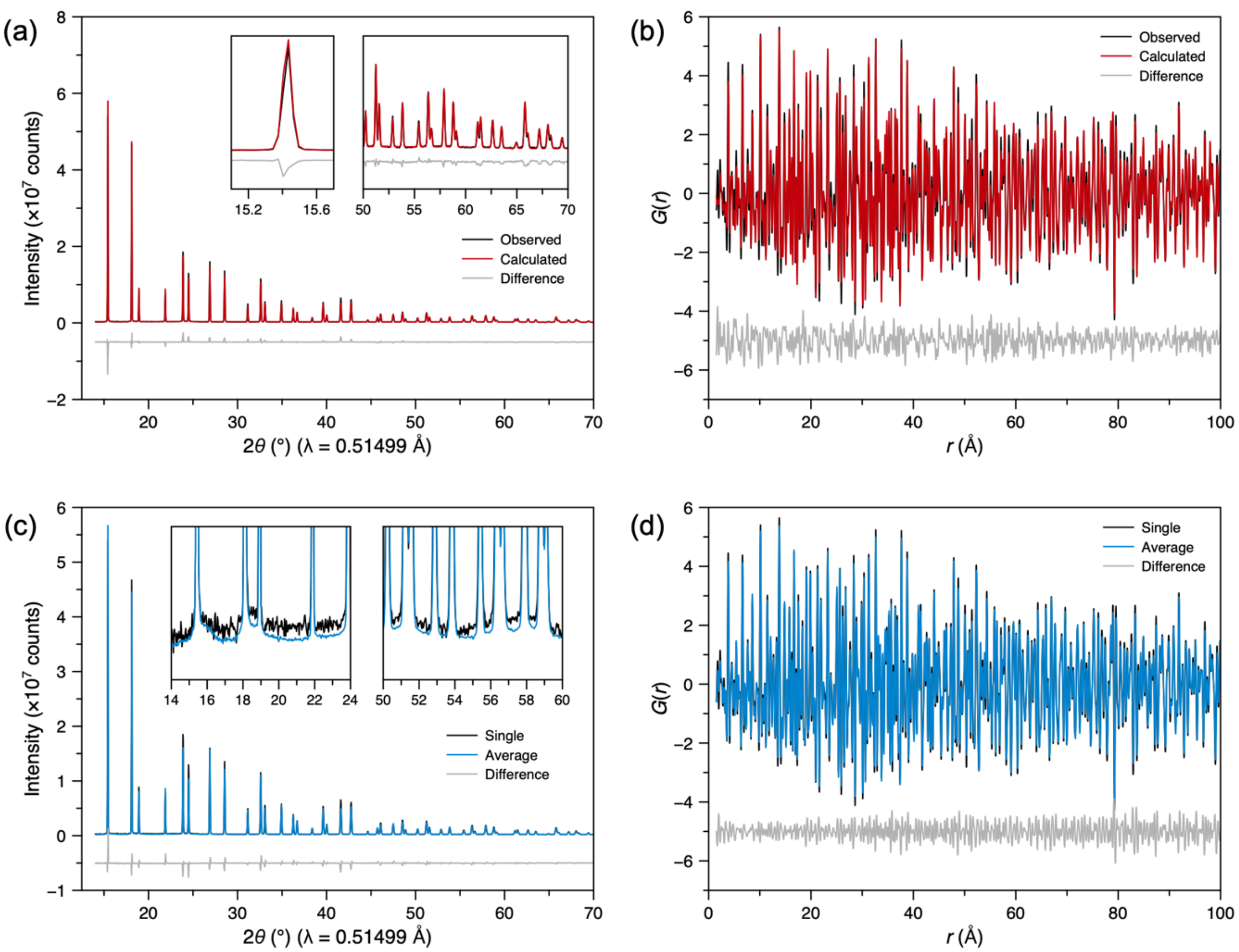

**Figure 5 (a)** Representative Rietveld refinement of normalised powder diffraction data of NIST $CeO_2$ 674 SRM in a fused silica capillary obtained from a single attenuated pulse of XFEL radiation. Insets show enlarged regions of the data at low- and high-2$\theta$. **(b)** Small-box 'real-space Rietveld' refinement against the single-pulse PDF obtained from the data in (a). **(c)** Comparison between diffraction data from a measurement using a single pulse and that averaged over 74 pulses. Insets show enlarged regions of the data at low- and high-2$\theta$. **(d)** shows the PDF data equivalent to the data in (c).

The $CeO_2$ single-pulse diffraction data were transformed into PDFs using PDFgetX3 (**Figure 5b**). Small-box refinement in TOPAS Academic yielded a mean $R_p$ of 18.1(1.5)%, $R_{wp}$ of 18.2(1.5)%, a lattice parameter of 5.4078(0.0017) Å, $B_{iso}$(Ce) of 0.165(0.022) and $B_{iso}$(O) of 1.58(0.26) Å$^2$ (mean and standard deviations from the same 74 single pulse datasets used for the Rietveld refinements given above). These are consistent with previously reported $B_{iso}$ values of 0.226 and 1.55 Å$^2$ for Ce and O, respectively; however, it is known that the values obtained from Rietveld refinement of the reciprocal space data are more reliable (Neder & Proffen, 2020). Overlays of multiple single-pulse reciprocal space data and PDFs revealed little variation between pulses, indicating high reproducibility (**Supplementary Figure 6**).

As expected, averaging multiple single-pulse measurements further improved the signal-to-noise (**Figures 5c & d**). However, even single-pulse data exhibited low noise, emphasising the exceptional quality of the setup. Rietveld and small-box refinements using the averaged data closely matched the single-pulse results (**Supplementary Figure 7**). The averaged Rietveld refinement yielded an $R_p$ of 4.846%, lattice parameter of 5.4075(5) Å, $B_{iso}$ (Ce) of 0.253(4) Å$^2$ and $B_{iso}$ (O) of 0.48(3) Å$^2$. The lattice parameter was almost identical to those obtained from the single-pulse measurements, emphasising the high reproducibility of our measurements. Again, these values compare well to previously reported results from synchrotron data (Yashima & Takizawa, 2010). The average small-box refinement yielded an $R_p$ of 16.022%, lattice parameter of 5.40747(4) Å, $B_{iso}$ (Ce) of 0.165(3) Å$^2$ and $B_{iso}$ (O) of 1.53(2) Å$^2$.

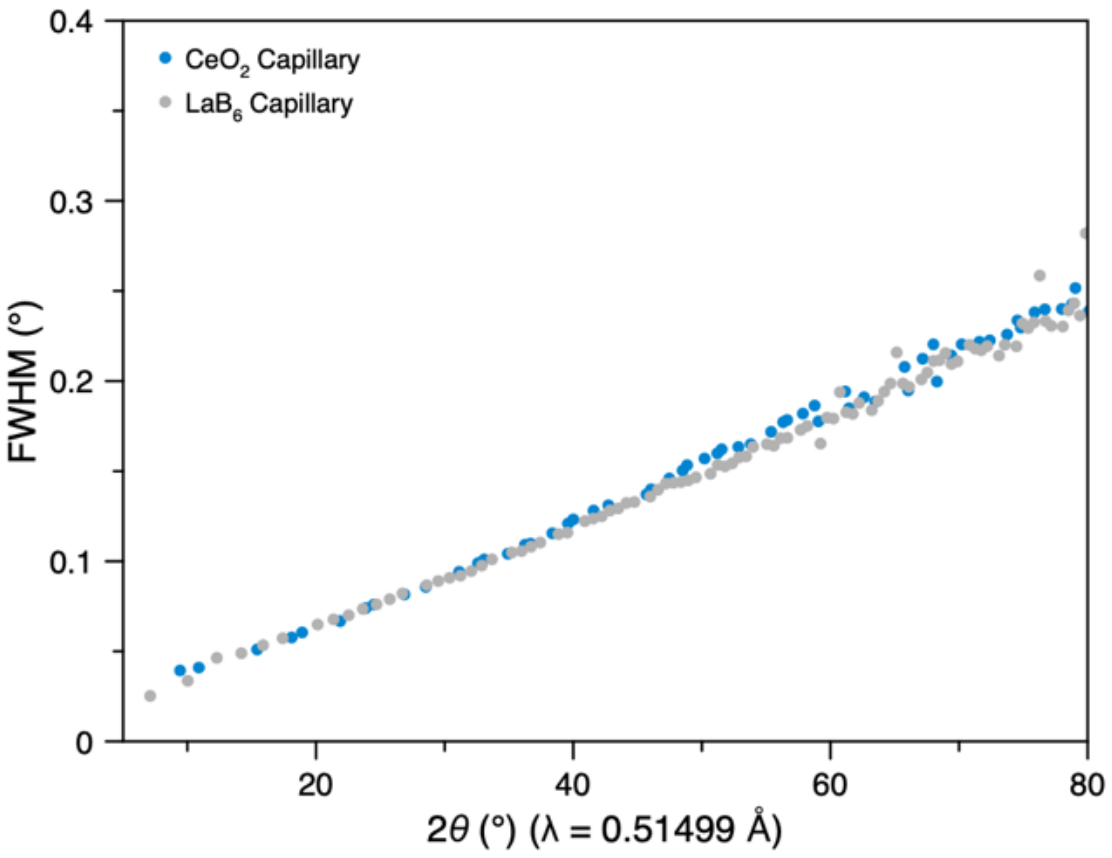

**Figure 6** Comparison between the Bragg peak full width at half maximum (FWHM) obtained from averaged powder diffraction data of $CeO_2$ and $LaB_6$ in capillaries, as a function of scattering angle, $2\theta$.

The instrumental resolution function (IRF) was extracted from the averaged $CeO_2$ data as a function of $2\theta$ (**Figure 6**). The variation in FWHM was around 0.03 to 0.25° in the range 10 to 80°. An essentially identical IRF was obtained from $LaB_6$ (**Figure 6**), with the peak broadening in both cases being dominated by instrumental contributions (**Supplementary Figure 8**).

*Crystalline standard for low-Q measurements: Ag behenate*

To assess performance at low $Q$-values, we measured silver behenate, known for its well-defined Bragg peaks extending down to $Q \sim 0.15$ Å$^{-1}$ (Blanton *et al.*, 2011). The expected low-$Q$ peaks were observed in the Jungfrau detector. The Bragg peaks were symmetric and well modelled using the Pseudo-Voigt peak shape. A Pawley refinement was successfully carried out using previously reported lattice parameters, confirming the setup's capability to cover both low- and high-$Q$ scattering (**Figure 7**).

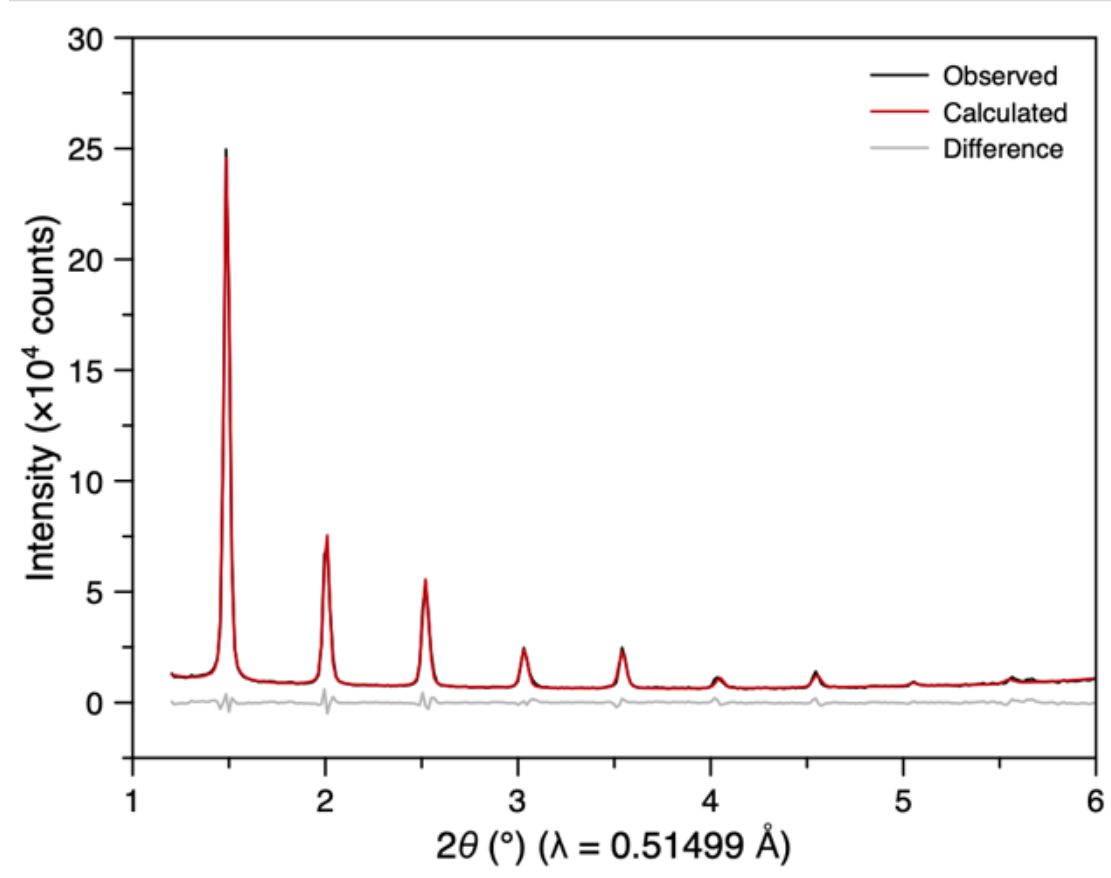

**Figure 7** Pawley refinement of silver behenate using averaged powder diffraction data from the low-angle Jungfrau detector.

*Nanocrystalline $TiO_2$ for total scattering analysis using DebUsSy*

A nanocrystalline $TiO_2$ P25 powder (with ~25 nm particles) was measured in capillary geometry. The data were averaged, corrected for absorption, and the signal from an empty quartz-glass capillary was subtracted (**Figure 8**). As expected, the Bragg peaks are broader than those seen in the Si and $CeO_2$ data due to the crystallite size of the $TiO_2$ powder. By making use of the IRF, the sample broadening from $TiO_2$ could then reliably be extracted and analysed.

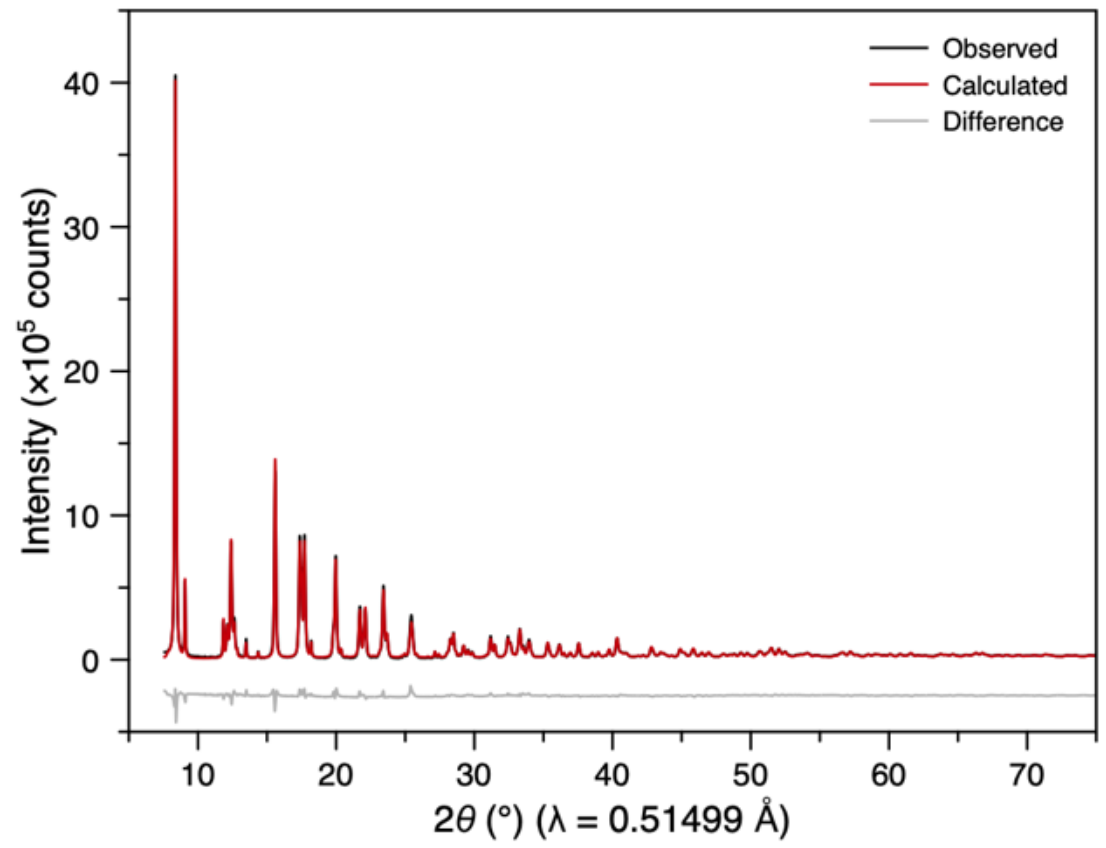

**Figure 8** Debye scattering analysis of averaged powder diffraction data from nanocrystalline $TiO_2$.

The sample contained ~90% anatase and ~10% rutile polymorphs of $TiO_2$. Reciprocal space data analysis was performed using the Debye scattering equation (within the DebUsSy program suite (Cervellino *et al.*, 2015)) and an atomistic model of nanocrystals similar to that described in (Bertolotti *et al.*, 2020). The anatase unit cell (refined from synchrotron data $a$ = $b$ = 3.7860 Å, $c$ = 9.5080 Å (Bertolotti *et al.*, 2020)) was used as a building block to generate a bivariate population of prismatic nanocrystals according to two independent growth directions, one along the $c$ axis ($L_c$) and the other in the $ab$ plane ($D_{ab}$). The minority rutile phase (composed of larger ∼50 nm nanocrystals) was modelled as a blank trace using the calculated pattern from TOPAS software (Coelho, 2018). A good fit ($R_p$ = 5.29%) was obtained over the range $1.6 < Q < 14.9$ Å$^{-1}$ with $B_{iso}$(Ti) = 0.20 Å$^2$ and $B_{iso}$(O) = 0.35 Å$^2$. These values are underestimated compared to those extracted from synchrotron data (collected at the X04SA Material Science beamline of the Swiss Light Source). By convoluting the instrumental peak broadening obtained from the IRF previously extracted from $LaB_6$ in **Figure 6** to the DSE calculated pattern, the sizes and their relative dispersions (σ) of the slightly anisotropic anatase nanocrystals were extracted ($D_{ab}$ = 23.88 nm, σ/ $D_{ab}$ = 0.37, $L_c$ = 17.91 nm, σ/ $L_c$ = 0.41), described by a discrete bivariate lognormal distribution function. These results are in excellent agreement with the previous results obtained from synchrotron data (Bertolotti *et al.*, 2020).

*Total scattering data from amorphous silica and a metallic glass: comparison with synchrotron data*

The structures of amorphous materials, lacking long-range order, can only be studied using total scattering/PDF methods and therefore these XFEL developments are especially important for this class of materials. PDFs for the amorphous samples reported here were generated using the GudrunX software, since this software produces absolutely normalised total scattering structure factors, $S(Q)$, and PDFs, $D(r)$[1], that can be directly compared with previously normalised data (Soper, 2011; Keen, 2001). The PDFs presented here are the sine Fourier transforms of $S(Q)$ using $Q$-values between 0.35 to 15 Å$^{-1}$.

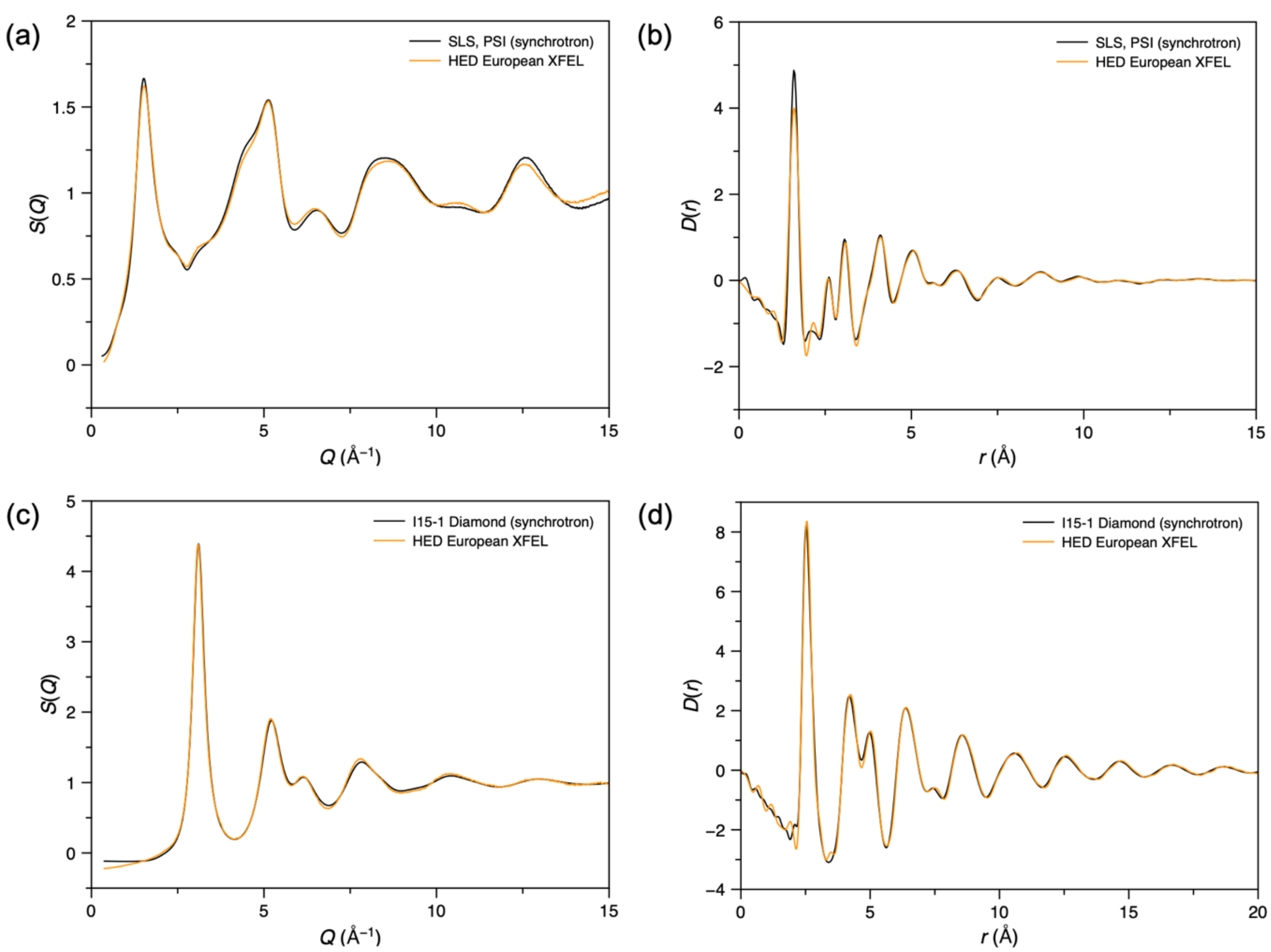

**Figure 9** Averaged $S(Q)$ and PDF data (**(a)** and **(b)**, respectively) for $SiO_2$ glass compared to similar synchrotron data (SLS, PSI, Switzerland). Synchrotron data were collected from a 2 mm diameter silica glass rod using a Mythen II detector and a 28 keV X-ray beam (unpublished data collected by Antonio Cervellino). Averaged $S(Q)$ and PDF data (**(c)** and **(d)**, respectively) for a single 25 μm layer of commercial metallic glass, $Fe_{78}B_{13}Si_9$, compared to similar data obtained from four layers of the same glass measured on the I15-1 diffractometer at Diamond Light Source (data measured by Daniel Irving as part of the I15-1 mail-in service, proposal CY39017).

For $SiO_2$ glass, the so-called first sharp diffraction peak (FSDP) was observed at 1.54 Å$^{-1}$ (**Figure 9a** & **Supplementary Figure 9**), consistent with previous reports (*e.g.* (Kohara & Suzuya, 2003)). The PDF contained the expected peaks corresponding to the $SiO_4$ tetrahedra in the structure (**Figure 9b**), with peaks at 1.6 (Si–O), 2.6 (O–O) and 3.1 Å (Si–Si). Comparisons with synchrotron data reveal remarkable agreement for the $S(Q)$ over the whole measured $Q$-range and for the PDF (**Figures 9a & b**, respectively). Moreover, there was only a 1.52% variation in the relative Si–O coordination numbers determined from the peak at 1.6 Å. These results highlight the advantage of the combined Varex–Jungfrau detector setup, which provides access to the low- and high-$Q$ values necessary for studying amorphous samples.

[1] Note on correlation functions: To avoid confusion, PDFs produced using PDFgetX3 or GudrunX are labelled '$G(r)$' or '$D(r)$', respectively. The '$G(r)$' function used here is the function '$G^{PDF}(r)$' in (Keen, 2001), which also describes how '$G^{PDF}(r)$' and '$D(r)$' are related.

A metallic glass ($Fe_{78}B_{13}Si_9$) was also measured, with excellent agreement between its $S(Q)$ and PDF and those measured on I15-1 at Diamond Light Source (**Figure 9c & 9d**). The most significant deviation between the two datasets is in the lowest-$Q$ region of $S(Q)$, where it is often difficult to get reproducible results between instruments. The first three low-$r$ peaks in the PDF are observed at 2.56, 4.22 and 5.02 Å (**Figure 9d**), most likely corresponding to the nearest, next-nearest and next-next-nearest Fe-Fe distances of an approximately close-packed arrangement of the iron atoms in the glass. Impressively, good quality $S(Q)$ and $D(r)$ data were successfully obtained from just a single XFEL pulse (~30 fs) (**Supplementary Figure 10**). Additionally, the $S(Q)$ and PDFs generated by GudrunX and PDFgetX3 were very similar (**Supplementary Figure 10**), except for some low-$r$ deviations in the PDF and an overall scale factor, due to the assumptions implicit in PDFgetX3's *ad-hoc* approach (Juhás *et al.*, 2013).

*Assessment of liquid and solution diffraction: water and Keggin solution*

In addition to amorphous solids, liquids are a critical class of materials studied using total scattering. Water was selected due to the extensive literature on its total scattering and for the challenge of its relatively weak X-ray scattering signal. A well-normalised $S(Q)$ was obtained using GudrunX, with the key double peak maxima observed at 2.08 and 2.93 $Å^{-1}$ (**Figure 10**), closely matching reference data (Soper, 2013). The PDF was generated, exhibiting the characteristic O–O peak present at 2.83 Å (**Supplementary Figure 11**), and was also in agreement with reference data (Soper, 2013).

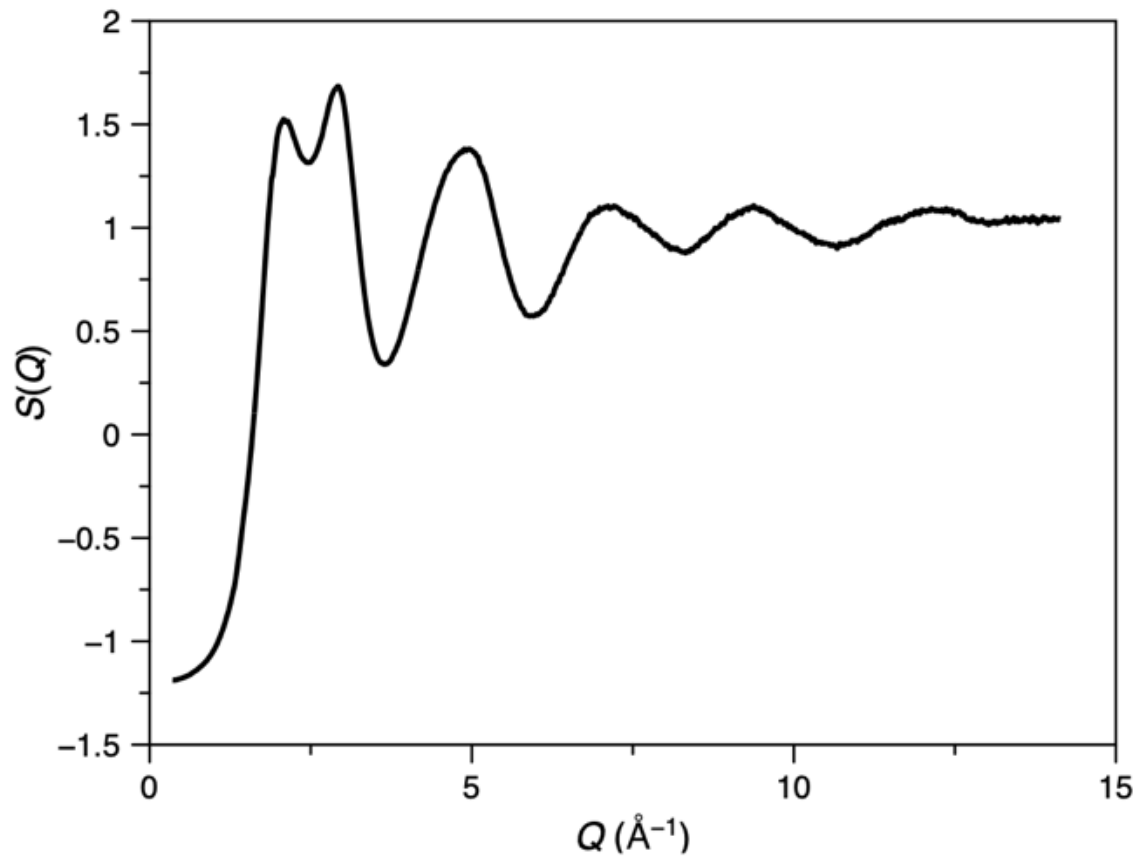

**Figure 10** Average *S(Q)* obtained from water.

We then tested the ability to detect atomically precise clusters in solution by studying the tungsten Keggin cluster ($[W_{12}O_{40}]^{6-}$), which forms upon dissolution of ammonium metatungstate in water. A 1.0 M solution (with respect to [W]) was loaded into a fused silica capillary and measured. As expected for a cluster in solution, over 70% of the total scattered intensity originated from the capillary and solvent in the $Q$-range 0.9 to 16.5 $Å^{-1}$ (**Figure 11a**). Nevertheless, a distinct signal from the Keggin structure was observed after subtracting the background from the capillary and solvent and was used to generate the PDF (**Figure 11b**). Key peaks in the PDF were observed at 1.91 Å (W–O) and 3.27 (W–W), 3.72 (W–W) and 6.05 Å (W–W). To further evaluate the sensitivity of the setup, we measured the Keggin cluster across a concentration range from 0.2 to 1.0 M (**Supplementary Figure 12a**). Even at the lowest concentration, a clear signal from the cluster was detected, and almost identical PDFs were generated across the range of concentrations.

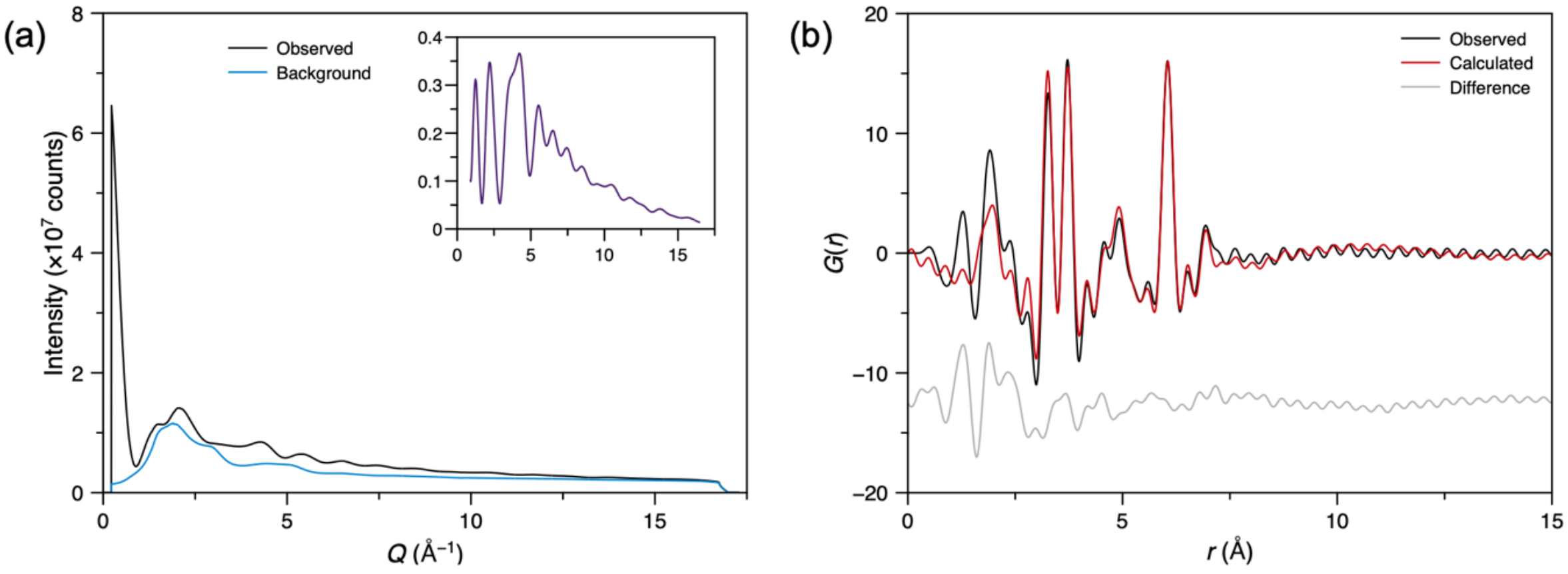

**Figure 11 (a)** Measured average intensity of the fused silica capillary loaded with the aqueous 1.0 M Keggin solution and a capillary loaded with water (background). Inset shows the sample intensity after background correction (omitting the intensity at low-$Q$ that most likely is a small-angle X-ray scattering contribution). **(b)** Average PDF of the Keggin cluster in solution and a model refined in DiffPy-CMI.

The structure of the Keggin cluster was then refined against the PDF using cluster modelling within the DiffPy-CMI framework (**Figure 11b**) (Juhás *et al.*, 2015). The atomic structure of the Keggin cluster was extracted from the ammonium metatungstate crystal structure (Magnard *et al.*, 2023). The PDF was then calculated from this structure using the Debye equation. To refine the model against the data, an overall scale factor, two isotropic thermal displacement parameters (*i.e.* for W and O), a parameter $\delta_2$ (which accounts for correlated motion effects), and a zoom-scale parameter (which stretches/compresses the atomic structure isotropically whilst maintaining relative atomic positions) were all varied in a similar approach to (Magnard *et al.*, 2023). A good refinement was obtained ($R_w$ = 33.8%), similar to previous reports (Magnard *et al.*, 2023), with a zoom-scale parameter of almost unity (1.01) and $B_{iso}$(W) = 0.22(2) Å$^2$ and $B_{iso}$(O) = 0.18(2) Å$^2$. Below $r = 2.5$ Å, the difference function exhibited more significant and structured features; these were identified as contributions from the ammonium cation and its solvent-restructuring effects, which were not accounted for in the model, by comparison to reference data collected at DanMAX, MAX IV (unpublished data collected by Adam Sapnik) (**Supplementary Figure 12b**).

As a final demonstration of the high-quality nature of the data, the PDF of the Keggin cluster was extracted from an $S(Q)$ of a 1 M Keggin solution measured using only a single XFEL pulse (~30 fs) over the $Q$-range of 0.9 to 16 Å$^{-1}$. This was then compared to a synchrotron measurement of 5 minutes from a 2 M solution collected at DanMAX, MAX IV (unpublished data collected by Adam Sapnik) (**Figure 12**). Despite more than 70% of the signal arising from background scattering, an increased concentration of the Keggin in the synchrotron measurement, and, most impressively, a 16 order of magnitude reduction in acquisition time, the PDF obtained from HED is of remarkable quality and is very clearly comparable to that obtained using synchrotron radiation. Minor differences in the PDFs are observed below 3 Å, where PDFs are typically less reliable and from differences in the counterion solvation environment with concentration. Beyond 10 Å, no structural signal is present, and the variations here arise from termination effects and the different signal-to-noise statistics of each measurement. Ultimately, this result alone highlights the success of our efforts, demonstrating high-quality ultra-fast PDF on one of our most challenging samples.

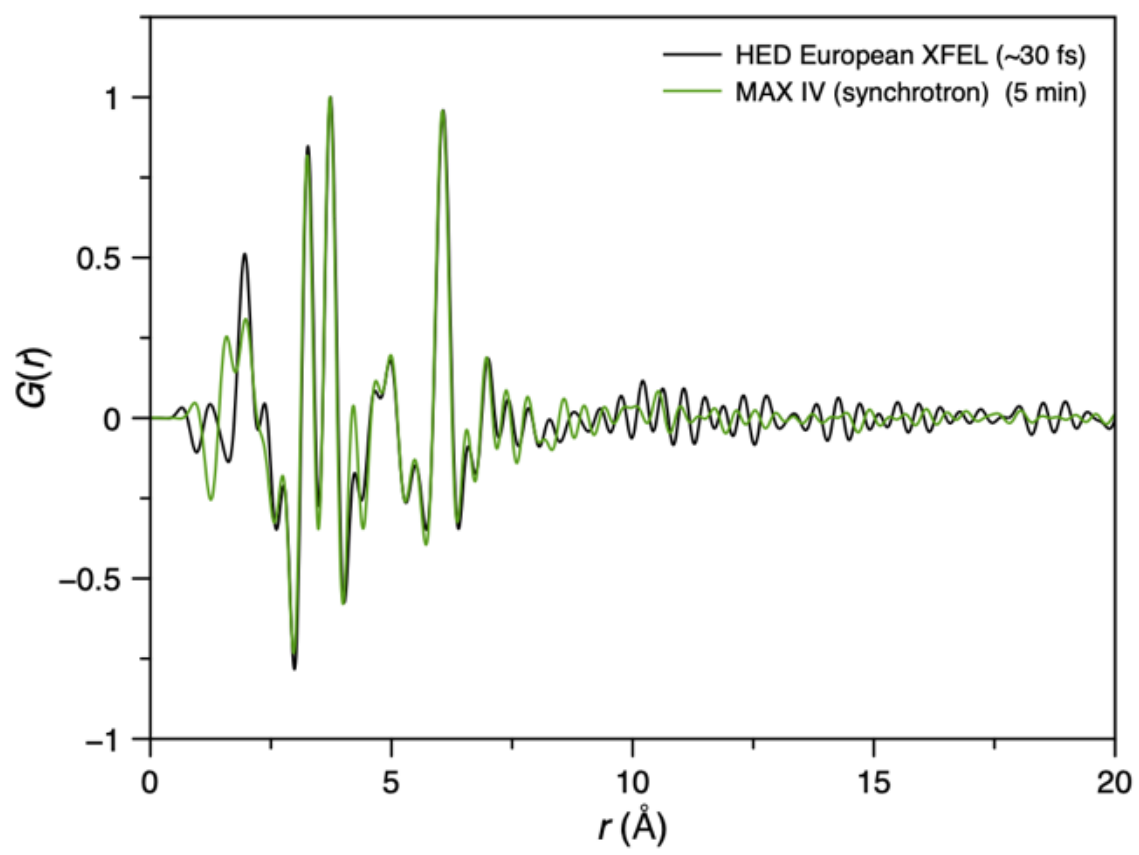

**Figure 12** Comparison between a PDF from 1.0M Keggin solution obtained using a single XFEL pulse (green) and a 5-minute synchrotron measurement of a 2.0 M Keggin solution (black).

## Conclusions and Outlook

Over the past forty years, total scattering and pair distribution function measurements have become increasingly crucial for determining short-range correlations within all types of condensed matter (crystalline, glass, liquid, *etc.*) (Keen, 2020). At the same time, relevant technological topics are increasingly reliant on the behaviour of local structures to understand materials' function. Examples of this include ionic conduction in battery materials, local distortions in ferroelectrics, or gas absorption in porous materials. If this is combined with the need to understand glass structure for increasingly 'smart' amorphous materials or the underlying fundamental physics of geologically relevant liquid structures under highly non-ambient conditions, or the nanoscale structure of non-equilibrium metastable quantum states exhibiting ultrafast optical or electrical switching, then it is clear that these techniques will continue to be important for some time to come.

It is important to note that many disordering processes—or the formation of novel nanoscale states—occur on timescales inaccessible to total scattering measurements at conventional synchrotron sources. These ultrafast dynamics can only be captured through carefully designed experiments at XFEL facilities, where the combination of femtosecond pulses and high photon energies provides unique access to this regime. Such measurements inherently require meticulous preparation and precision but are essential for probing processes that are far too rapid for even the most advanced synchrotron techniques. The outlook is highly promising—strengthened by the results presented in this work and the continued advancement toward higher-energy X-rays at XFEL facilities. We are confident that total scattering and PDF measurements at XFELs will steadily improve, offering increasingly profound insights into the structural dynamics of complex materials.

We have shown that quantitative total scattering data can be obtained from carefully designed XFEL instrumentation over a wide range of $Q$-values, achieving a $Q_{max}$ above 16 $Å^{-1}$ whilst simultaneously measuring data down to a $Q_{min}$ of 0.35 $Å^{-1}$. This maximum $Q$ is nearly twice what is currently routine and a third higher than has been achieved for the less stringent $\Delta S(Q)$ measurements often used for pump–probe experiments. The data are now sufficient to produce PDFs with reasonable real-space resolution for quantitative studies of disordering structural processes. We have also shown that higher $Q_{max}$ are possible – to values approaching 20 $Å^{-1}$ (see **Figures A1–4** in **Appendix I**) – but these are currently only possible by collecting data at scattering angles above $2\theta = 90°$ which starts to negate the benefits of the angled Varex detector and compromises the versatility of having sample mounting systems perpendicular to the incident X-ray beam. Routine total scattering measurements with $Q_{max}$ greater than 16 Å$^{-1}$ will, therefore, probably only be realistic following XFEL source upgrades that provide higher energy X-ray beams.

The key result of this work is that high-quality, quantitative total scattering and PDFs have been obtained from single ~30 fs X-ray pulses. We have demonstrated this using a range of samples with a wide variety of structure types, all of which have produced data of similar quality to that available at synchrotron sources over the range of $Q$ that we have measured. The experiments have been carried out on the HED instrument at the European XFEL, but the specific apparatus needed for the measurements (Varex detector, background-reducing components, *etc*.) is readily available and could be installed at other high-energy XFEL instruments. We have also developed data analysis pipelines to simplify the data collection and normalisation process, and we have shown that these produce data suitable for most of the data normalisation and analysis tools currently in use by the community.

We are confident that this work will pave the way for new time-resolved studies in catalysis, battery materials, phase transitions, nanoparticle nucleation, nanoscale structure characterisation of hidden non-equilibrium states, and any area of science where time-resolved total scattering or PDFs will be of benefit. Hence, the next obvious step is to link these measurements to pump–probe techniques where delays of tens of femtoseconds and longer can be used to study (*e.g.*) optical-electronic-phononic induced changes. In this regard, the positioning of the detectors is helpful because they do not obscure the horizontal plane or substantially obscure the vertical plane perpendicular to the X-ray beam, both of which pass through the sample position. Hence pump and diagnostic lasers can be used in their typical (*i.e.* horizontal) arrangements, as well as vertical applied fields and liquid-jet sample delivery systems.  The quality of PDF data available from solutions shows that we have an important tool for studies of rapid conformational changes in molecules in solution. The measurements are also highly relevant for those studying high-pressure, high-temperature liquids produced by shock waves. The high-pressure shock wave community is strong, and the quality of the total scattering data now available means that measurements of more complex compounds, where higher-resolution data are essential, are now tractable.

## Data Availability Statement

Experiment archives are available at 10.22003/XFEL.EU-DATA-007316-00 (for data presented in the main text) and 10.22003/XFEL.EU-DATA-003248-00 (for data presented in Appendix I).

## Author Contributions

The contributions from each author are listed below: (A) conceptualisation, (B) methodology, (C) software, (D) data analysis and modelling, (E) investigation, (F) resources, (G) writing: original draft and (H) writing: review and editing.

**AFS** (ACDEFGH), **PAC** (ABCDEH), **DSK** (ABCDEH), **JSOE** (ACDEFH), **FB** (ACDEFH), **AG** (ADEFH), **LJS** (CDEFH), **EAH** (EDFH), **ABB** (CDEFH), **RSS** (DEFH), **AD** (BEF), **CP** (C), **BDK** (E), **AP** (E), **IU** (F), **KGM** (DF), **TAB** (C), **VK** (EFH), **WL** (E), **ALG** (ABEF), **BBI** (ABFH), **CC** (EF), **ESB** (EH), **KMØJ** (ABEFH), **EEM** (BEF), **RBN** (ABDEF), **IR** (AB), **JSW** (AB), **MA** (BEH), **UB** (A), **EB** (CE), **CC** (E), **VC** (BH), **SG** (BE), **HH** (BE), **OSH** (BCE), **ZK** (E), **NK** (E), **TM** (C), **MN** (BEH), **AP** (B), **TRP** (ABEH), **LR** (E), **MR** (B), **AS** (B), **CS** (BEF), **MT** (E), **PT** (B), **UZ** (B), **KA** (ABDEFH), **DAK** (ABDEFGH).

## Acknowledgements

We acknowledge European XFEL in Schenefeld, Germany, for the provision of XFEL beamtime at Scientific Instrument HED (High Energy Density Science) under proposals #3248 and #7136. I15-1 data in Figure 9 were collected by Daniel Irving as part of the I15-1 mail-in service, proposal CY39017, Diamond Light Source. We are grateful to Antonio Cervellino for collecting the SLS (PSI, Switzerland) synchrotron data plotted in Figure 9 for DAK in 2010. We acknowledge DESY (Hamburg, Germany), a member of the Helmholtz Association HGF, for the provision of experimental facilities. Parts of this research were carried out at PETRA III and we would like to thank Martin Aaskov Karlsen for assistance in using beamline P02.1. Beamtime was allocated for proposal I-20230990 EC. The authors acknowledge MAX IV Laboratory for time on Beamline DanMAX under Proposal 20231682 and would like to thank Mads R.V. Jørgensen and Frederik H. Gjørup for their assistance. Research conducted at MAX IV is supported by the Swedish Research Council under contract 2018-07152, the Swedish Governmental Agency for Innovation Systems under contract 2018-04969 and Formas under contract 2019-02496. DanMAX is funded by the NUFI grant no. 4059-00009B. We thank Stefan Kycia (University of Guelph) for useful discussions about the inclined Varex detector arrangement. We thank Theo Maltezopoulos for his work enabling the XGM to operate at high photon energies and Frank Brinker for making lasing available at 24 keV. The authors are indebted to the HIBEF UC for the provision of instrumentation and Staff that enabled this experiment.

## Funding

AFS, RSS and KMØJ thank the Villum Foundation (42079), the Danish National Research Foundation (DNRF149) and the Novo Nordisk Foundation (0085640) for funding. The authors also thank the Danish Agency for Science, Technology, and Innovation for funding the instrument centre DanScatt. LJS, ABB, VAK, and BBI thank the Villum Foundation (25861). EAH thanks the UK Hub for the Physical Sciences on XFELS (HPSX) for part-funding her DPhil Studentship. CC (Crepisson) acknowledges support from the UK EPSRC under Grant No. EP/W010097/1. MF acknowledges support from the UK EPSRC under Grant No. EP/X025373/1 and Amplifi PP. ESB supported by the European Union's Horizon Europe research and innovation programme under grant agreement No 101185375. This work was supported by the UK Research & Innovation Future Leaders Fellowship (MR/W008211/1). JSW gratefully acknowledges support from EPSRC under research grant EP/X031624/1. TS appreciates support from AWE *via* the Oxford Centre for High Energy Density Science (OxCHEDS). CC (Camarda) acknowledges funding by the Deutsche Forschungsgemeinschaft (DFG, German Research Foundation) under contract 521549147 in grant AP262/3-1, DFG Priority program Deep Dyn SPP 2404.

# Appendix I

*Summary of HED experiment using four Jungfrau detectors around the sample.*

Our initial aim was to collect data to a maximum $Q$ of ~20 Å$^{-1}$. To do this, we placed four Jungfrau detectors around the sample position in IC1 of HED in the vertical plane (see **Figure A1**). The three within the vacuum chamber were above the horizontal and extended to a scattering angle of ~110°. There was also a detector outside the vacuum chamber of IC1 (see **Figure A1**) that was below the horizontal and covered some of the low-2$\theta$ angles not accessible to the lowest Jungfrau within the IC1 vacuum chamber. This arrangement gave a $Q_{max}$ of 19.9 Å$^{-1}$ but at the expense of gaps in 2$\theta$ coverage where there were no detectors. In order to provide access to these higher scattering angles, the sample stage was mounted at 120° to the horizontal (*i.e.* the normal to the plane of the stage pointed to a 2$\theta$ of 30°).

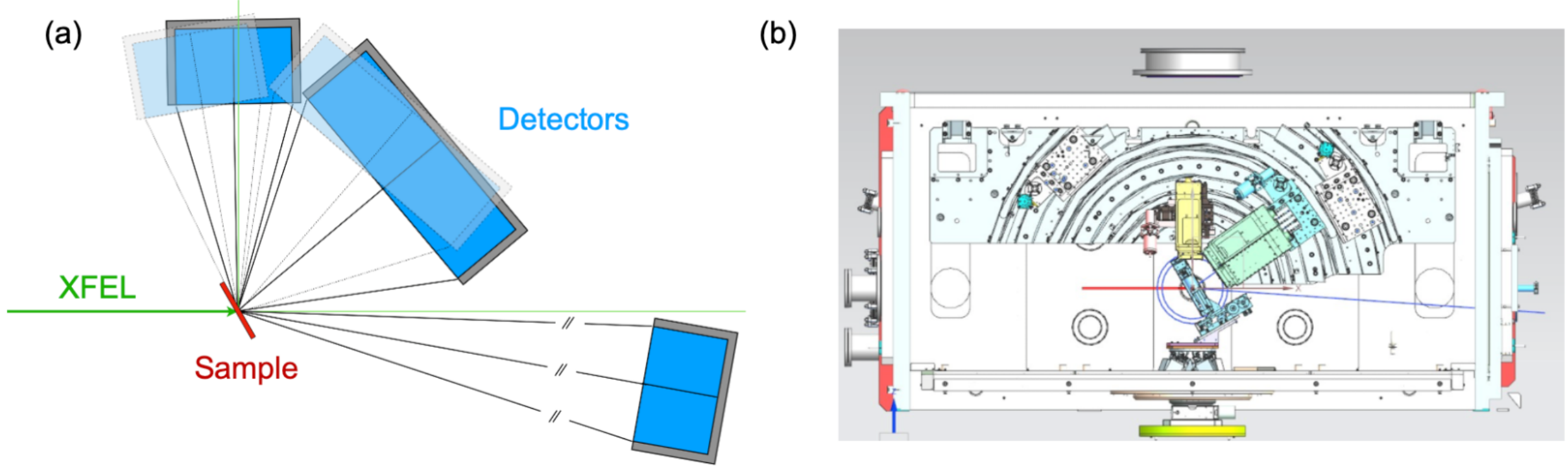

**Figure A1 (a)** Schematic and **(b)** CAD drawing of the detector arrangement using four Jungfrau detectors. The fourth detector placed below the X-ray beam outside the IC1 vacuum vessel is not shown in **(b),** and the three highest angle detectors were mounted on movable arms, although in practice they were not rotated.

Limitations on space and geometry meant that the $Q$-range was not continuous (see **Figure A2a**). We explored several options for interpolating the missing data between detectors with some success. For crystalline samples, where the Bragg scattering is orders of magnitude greater than the diffuse scattering, robust interpolation routines could be developed with little impact on the PDF (see **Figure A2b & A3a**). For samples where the signal is dominated by diffuse scattering, interpolation proved more challenging and severely impacted the PDF, giving rise to various artefacts. Furthermore, the data at high-$Q$ were fairly weak due to a low solid angle coverage in this region. Given our aim of establishing high-quality total scattering that is reliable, reproducible and robust for highly crystalline materials and a broad cross-section of materials (including liquids and glasses), we chose to pursue the Varex detector setup described in the main text instead.

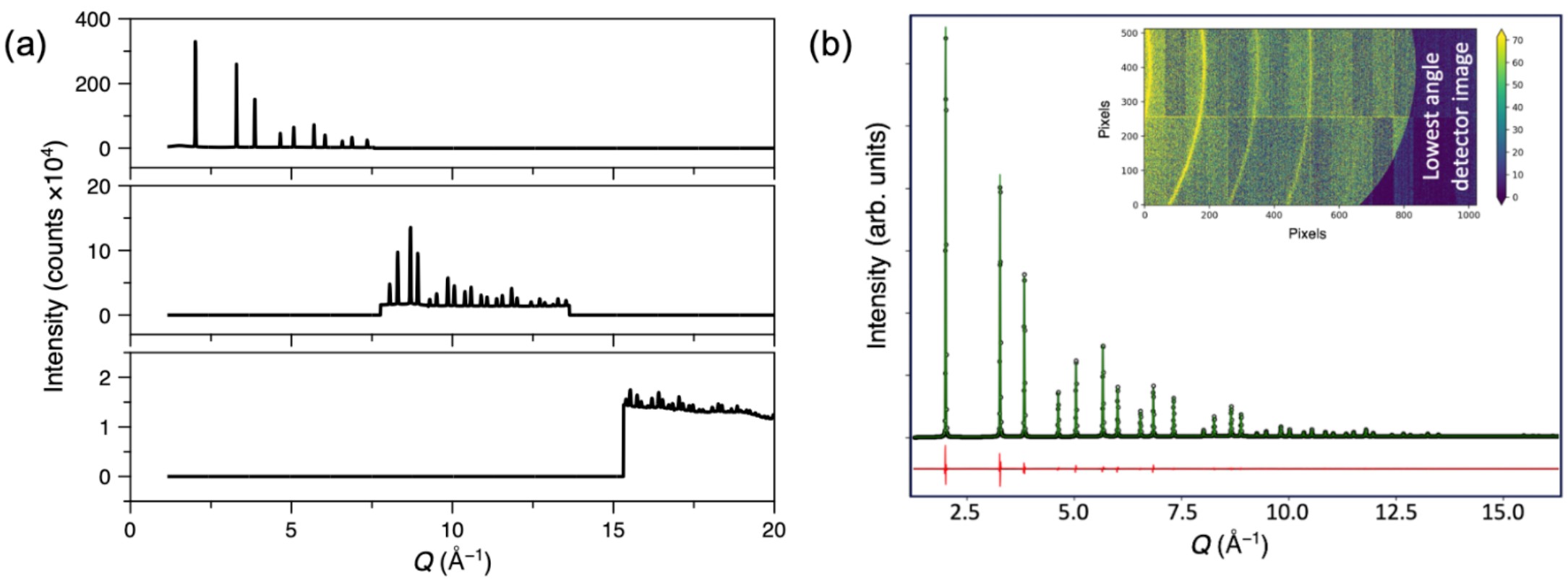

**Figure A2** Integrated data from a NIST Si 640b sample from the three highest-angle detectors shown **(a)** separately and **(b)** merged into a single diffraction pattern (points). (b) also shows a Rietveld refinement (green line), difference (red line, offset) and a 2D image of the data in the fourth (lowest angle) detector of Ag behenate showing its characteristic low-*Q* powder lines (inset).

Nonetheless, the PDFs that we extracted from some standard materials (see **Figures A3 & A4**) were reasonably robust and could be fitted with the calculated functions based on their published structures.

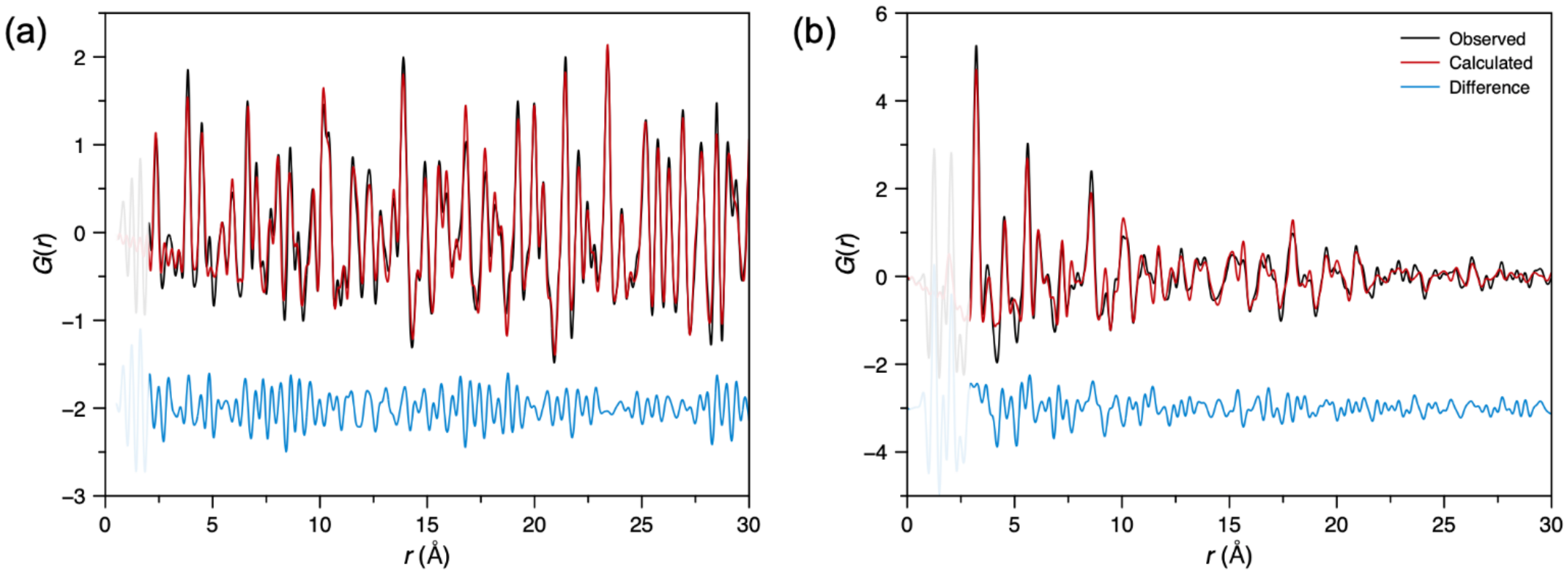

**Figure A3 (a)** Small-box refinement of NIST silicon 640b using average PDF data. **(b)** Refinement of ZnO nanoparticles against a PDF obtained from only a single pulse.

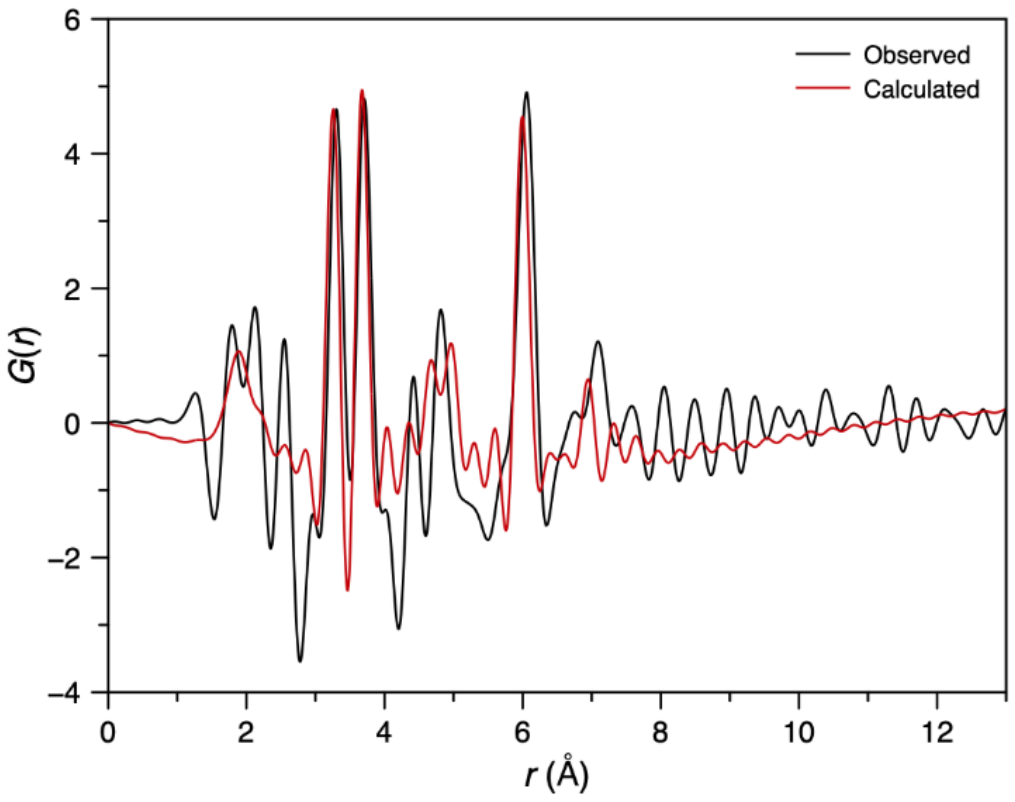

**Figure A4** Comparison between the experimental and calculated PDF from the Keggin cluster in water.

# High-Quality Ultra-Fast Total Scattering and Pair Distribution Function Data using an X-ray Free Electron Laser

## SUPPLEMENTARY INFORMATION

Adam F. Sapnik,[a]* Philip A. Chater,[b] Dean S. Keeble,[b] John S. O. Evans,[c] Federica Bertolotti,[d] Antonietta Guagliardi,[e] Lise J. Støckler,[f] Elodie A. Harbourne,[g] Anders B. Borup,[f] Rebecca S. Silberg,[a] Adrien Descamps,[h] Clemens Prescher,[i] Benjamin D. Klee,[j] Axel Phelipeau,[k] Imran Ullah,[l] Kárel G. Medina,[l] Tobias A. Bird,[b] Viktoria Kaznelson,[f] William Lynn,[h] Andrew L. Goodwin,[g] Bo B. Iversen,[f] Celine Crepisson,[m] Emil S. Bozin,[n] Kirsten M. Ø. Jensen,[a] Emma E. McBride,[h] Reinhard B. Neder,[l] Ian Robinson,[o] Justin Wark,[m] Michal Andrzejewski,[p] Ulrike Boesenberg,[p] Erik Brambrink,[p] Carolina Camarda,[p] Valerio Cerantola,[p,q] Sebastian Goede,[p] Hauke Höppner,[r] Oliver S. Humphries,[p] Zuzana Konopkova,[p] Naresh Kujala,[p] Thomas Michelat,[p] Motoaki Nakatsutsumi,[p] Alexander Pelka,[r] Thomas R. Preston,[p] Lisa Randolph,[p] Michael Roeper,[k] Andreas Schmidt,[p] Cornelius Strohm,[k] Minxue Tang,[p] Peter Talkovski,[k] Ulf Zastrau,[p] Karen Appel,[p] and David A. Keen.[s]*

**a**: Department of Chemistry, University of Copenhagen, Universitetsparken 5, 2100 Copenhagen Ø, Denmark.
**b**: Diamond Light Source Ltd, Diamond House, Harwell Science & Innovation Campus, Didcot, Oxfordshire, UK.
**c**: Department of Chemistry, University Science Site, Durham University, South Road, Durham, DH1 3LE, UK.
**d**: Dipartimento di Scienza e Alta Tecnologia and To.Sca.Lab, University of Insubria, Como, Italy.
**e**: Istituto di Cristallografia and To.Sca.Lab, CNR, Como, Italy.
**f**: Center for Integrated Materials Research, Department of Chemistry and iNANO, Aarhus University, Langelandsgade 140, 8000 Aarhus C, Denmark.
**g**: Department of Chemistry, Inorganic Chemistry Laboratory, University of Oxford, South Parks Road, Oxford OX1 3QR, UK.
**h**: School of Mathematics and Physics, Queen's University Belfast, University Road, Belfast BT7 1NN, UK.
**i**: Institute of Earth and Environmental Sciences, University of Freiburg, Freiburg, Germany.
**j**: HUN-REN Wigner Research Centre for Physics, Konkoly-Thege Miklós út 29-33, 1121 Budapest, Hungary.
**k**: Deutsches Elektronen-Synchrotron DESY, Hamburg, Germany.
**l**: Friedrich-Alexander Universität Erlangen–Nürnberg, Staudtstrasse 3, D-91058 Erlangen, Germany.
**m**: Department of Physics, Clarendon Laboratory, University of Oxford, Parks Road, Oxford OX1 3PU, UK.
**n**: Center for Solid State Physics and New Materials, Institute of Physics Belgrade, University of Belgrade, Pregrevica 118, 11080 Belgrade, Serbia.
**o**: London Centre for Nanotechnology, University College London, London WC1E 6BT, UK.
**p**: European XFEL, Holzkoppel 4, 22869 Schenefeld, Germany.
**q**: University of Milano-Bicocca, Piazza della Scienza 4, 20126, Italy.
**r**: Helmholtz-Zentrum Dresden-Rossendorf (HZDR), Dresden, Germany.
**s**: ISIS Facility, Rutherford Appleton Laboratory, Harwell Campus, Didcot, Oxfordshire OX11 0QX, UK.

Email: afs@chem.ku.dk & david.keen@stfc.ac.uk

# Supplementary Information

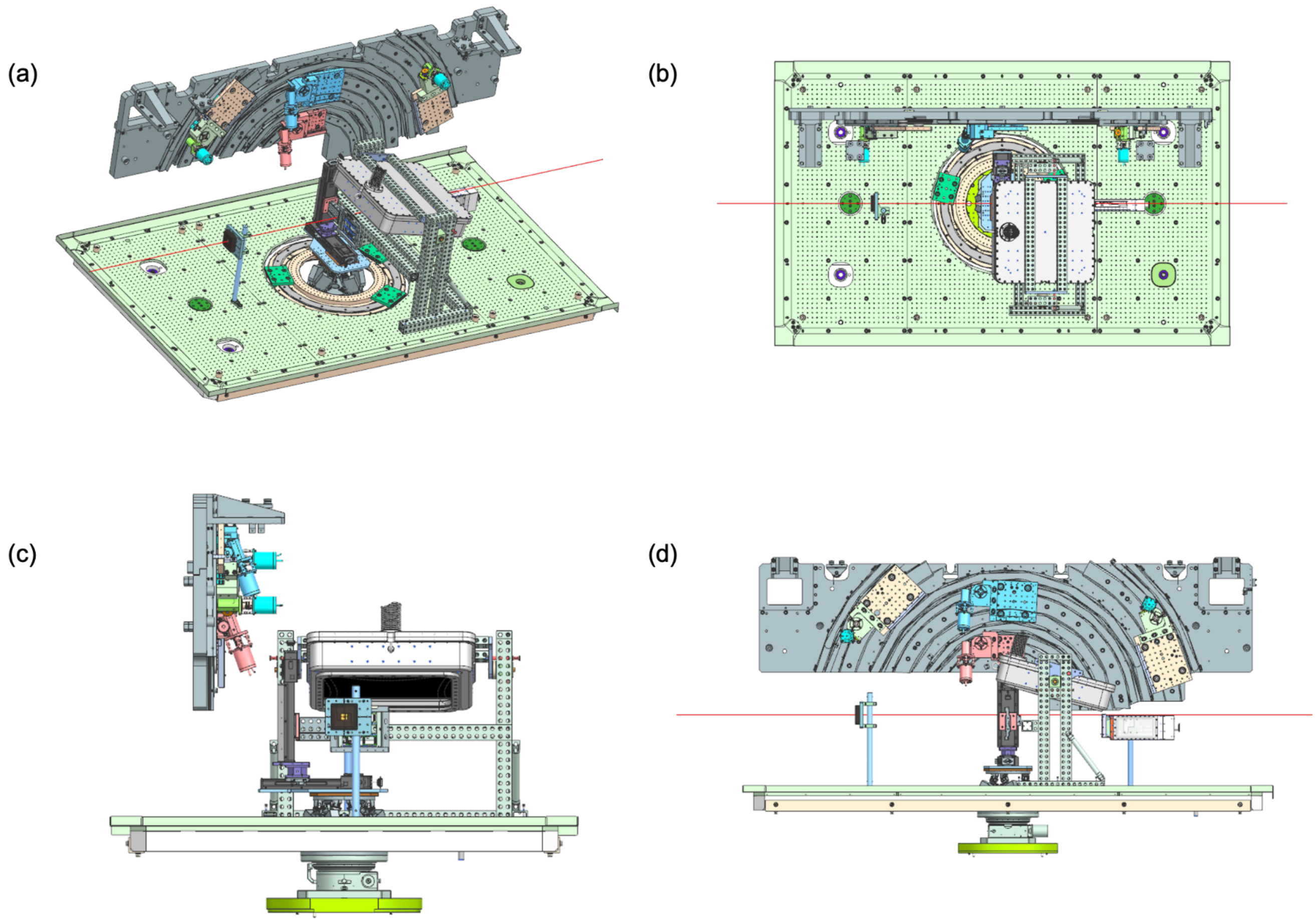

**Figure S1** CAD drawings of the HED setup, with the path of the XFEL beam shown in red.

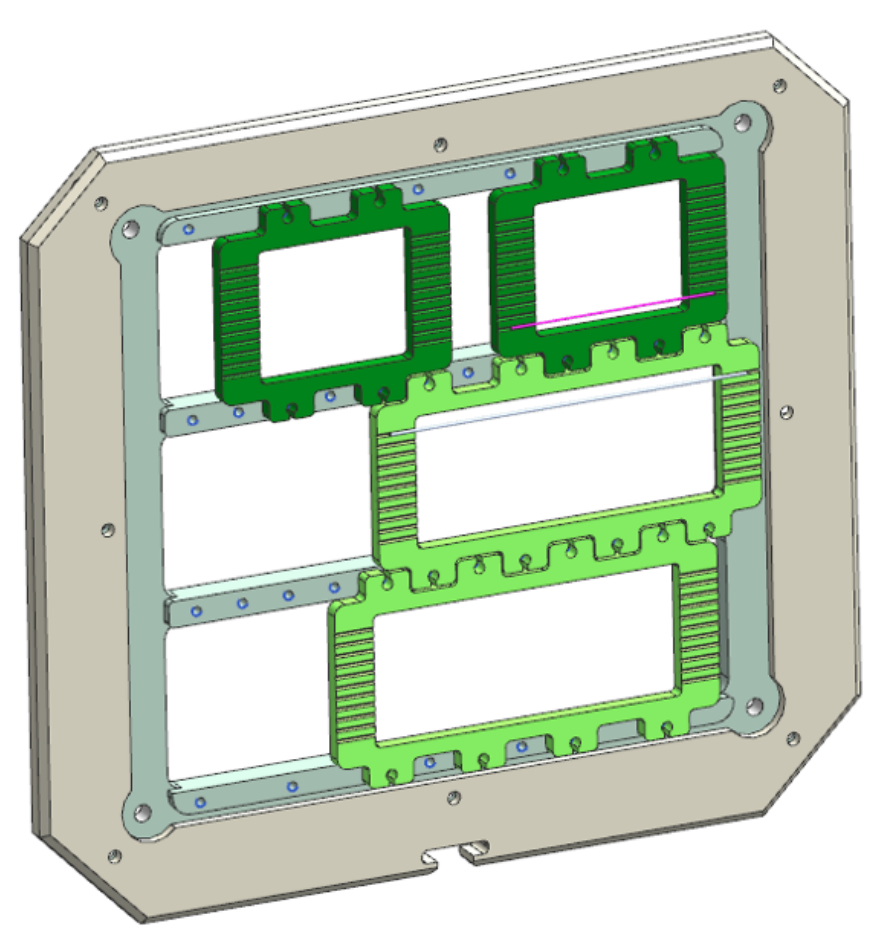

**Figure S2** EUCALL sample holder with bespoke (green in the diagram) frames that can accommodate long and short capillaries of varying diameter. Flat plates can also be mounted (not shown).

## Data Processing Corrections

### Sensor Efficiency correction

The sensors of the Varex and Jungfrau are 0.55 mm thick CsI and 0.45 mm thick Si, respectively. We extracted the cosine of the angle of incidence from the calibrated pyFAI azimuthalIntegrator object directly and used this angle to calculate an effective detector thickness (*d*), which was combined with the theoretical absorption length (μ) to give a pixel-wise sensor efficiency correction.

### Aluminium window transmission correction

We calculated the transmission of the aluminium windows based on the designed specifications and applied this to the detector images before azimuthal integration. Both detectors used a 0.4 mm-thick aluminium window. We extracted the cosine of the angle of incidence from the calibrated pyFAI azimuthalIntegrator object directly and inversely scaled the calculated −μd by this array to give a pixel-wise window transmission correction. It is known that the Varex window does not remain flat after it is subject to the pressure differences in the HED interaction chambers (Gorman *et al.*, 2024). Our observation was that this mostly impacts the edges of the window, which, in our detector arrangement, made little difference to the data normalisation except perhaps at the very highest scattering angles. Hence, our assumption of a flat window was sufficient.

### Detector Response

In the absence of a measured flat field from the supplier, we attempted to calculate a simple flat field correction to remove the most significant artefacts. The main features impacting our analyses were the "panelling" effect as seen in **Figure S3**. We then defined a flatfield correction consisting of 24 panels, each with a constant flat field value. We optimised the flat field by minimising the difference between neighbouring pixels across the panel boundaries.

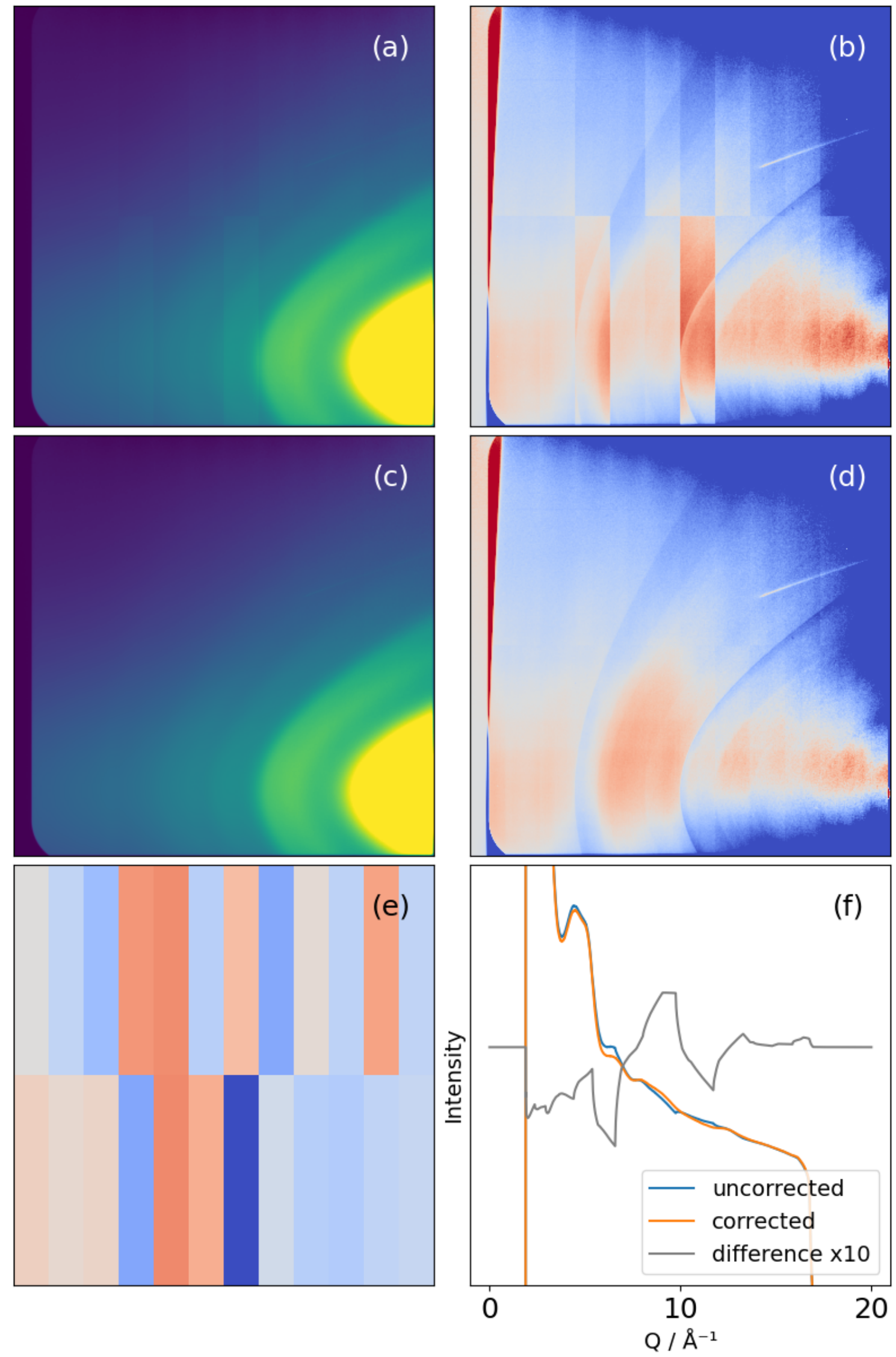

**Figure S3** The flat field correction, showing the artefacts causing the largest issues and our resolution. **(a)** An exemplar dataset from the Varex detector, **(b)** the difference between these data and the back-calculated images calculated from the median filtered data, to emphasise the panelling, **(c)** the raw data with the flat field applied, **(d)** the difference between the flat field corrected data and the back-calculated images calculated from the median filtered data, to emphasise the panelling **(e)** the flat field, **(f)** the impact of the flat field on the azimuthally integrated data.

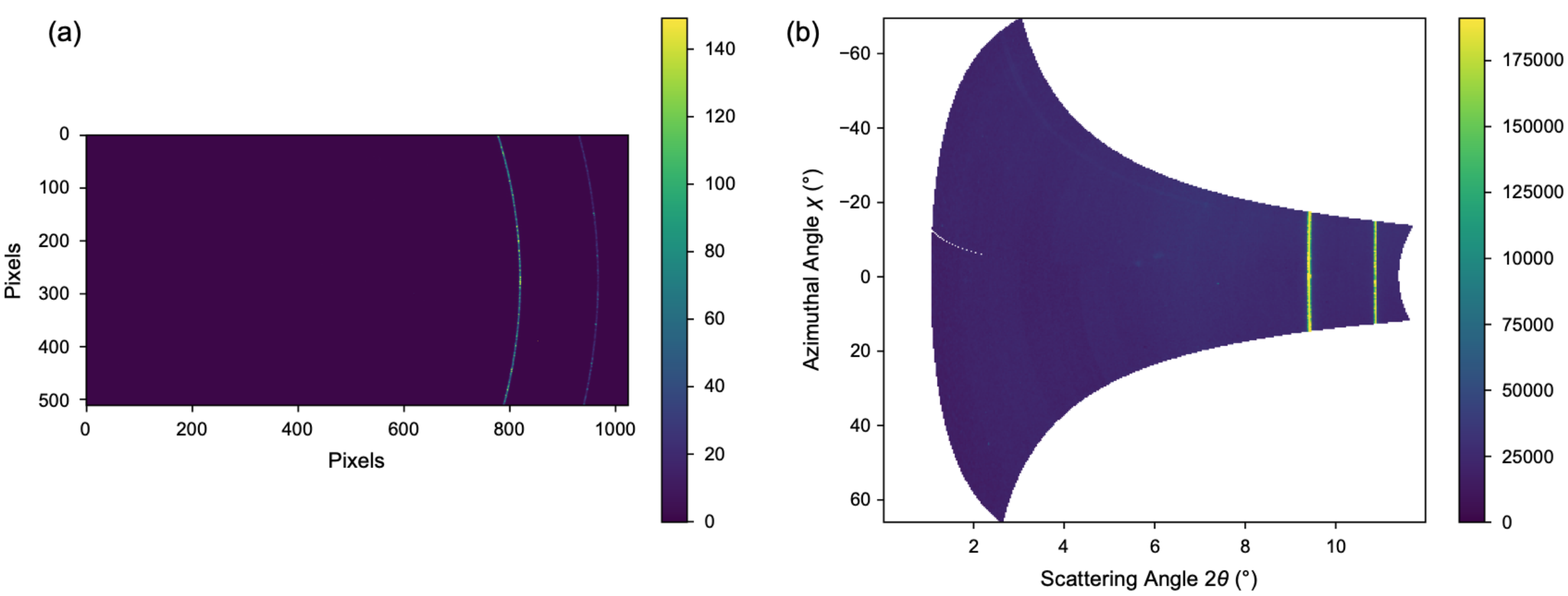

**Figure S4 (a)** Jungfrau detector image of a NIST $CeO_2$ 674 powder. **(b)** Data plotted as a function of scattering and azimuthal angle showing vertical lines of intensity corresponding to the two highest *d*-spacing Bragg reflections in NIST $CeO_2$ 674.

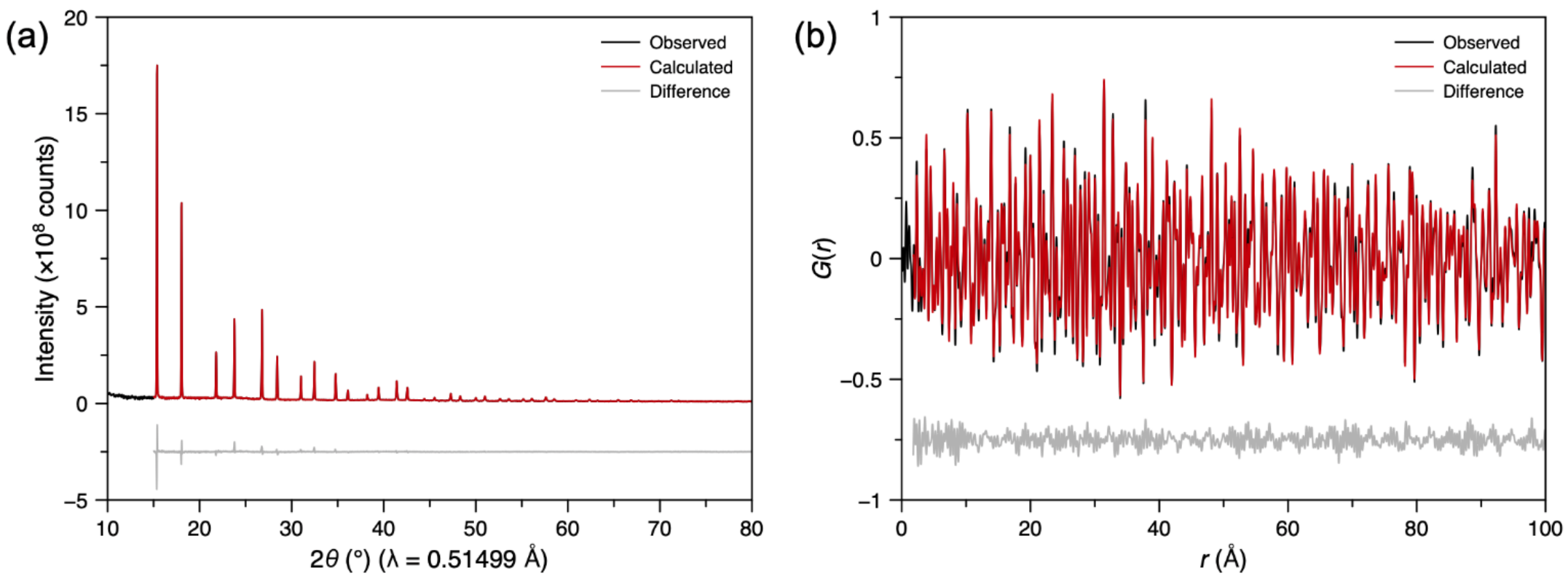

**Figure S5** Si NIST 640b **(a)** Rietveld refinement against the powder diffraction pattern and **(b)** small-box refinement against the PDF. Each of the two datasets were used to refine the structural model independently.

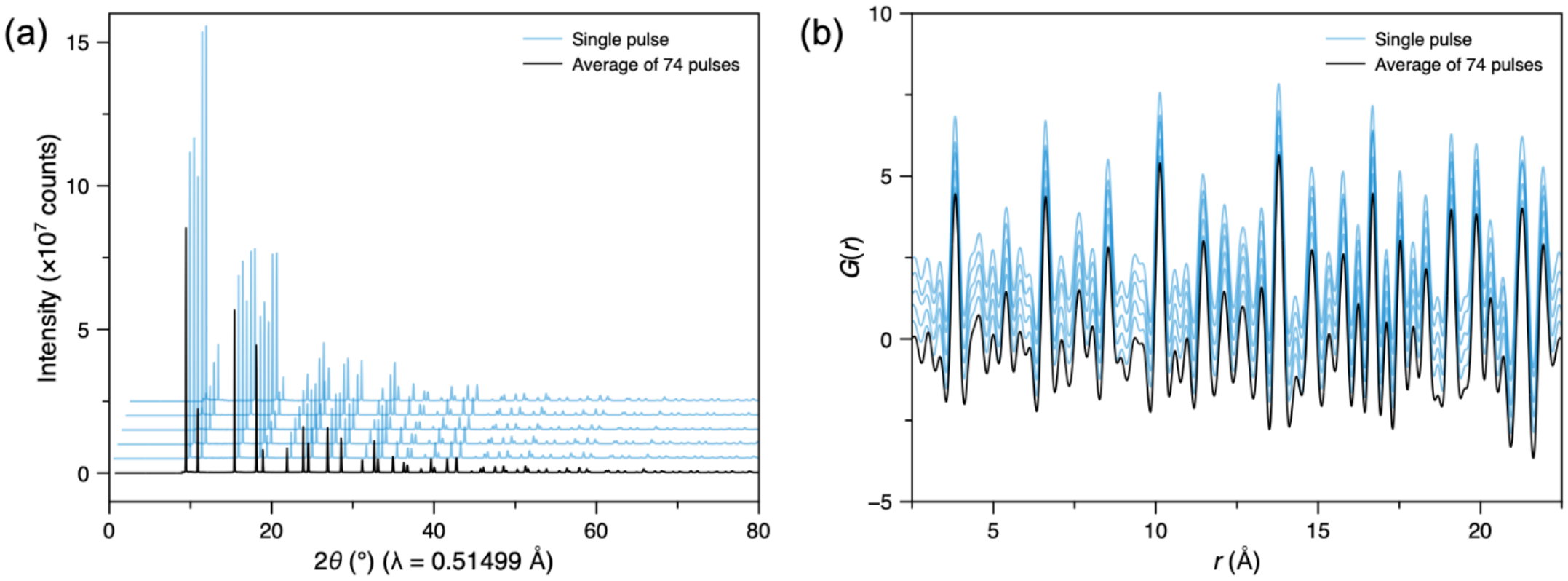

**Figure S6** Representative single pulse data (blue lines) compared with the average data (black line) for NIST $CeO_2$ 674. **(a)** Powder diffraction pattern (offset both horizontally and vertically for clarity) and **(b)** PDF (offset vertically).

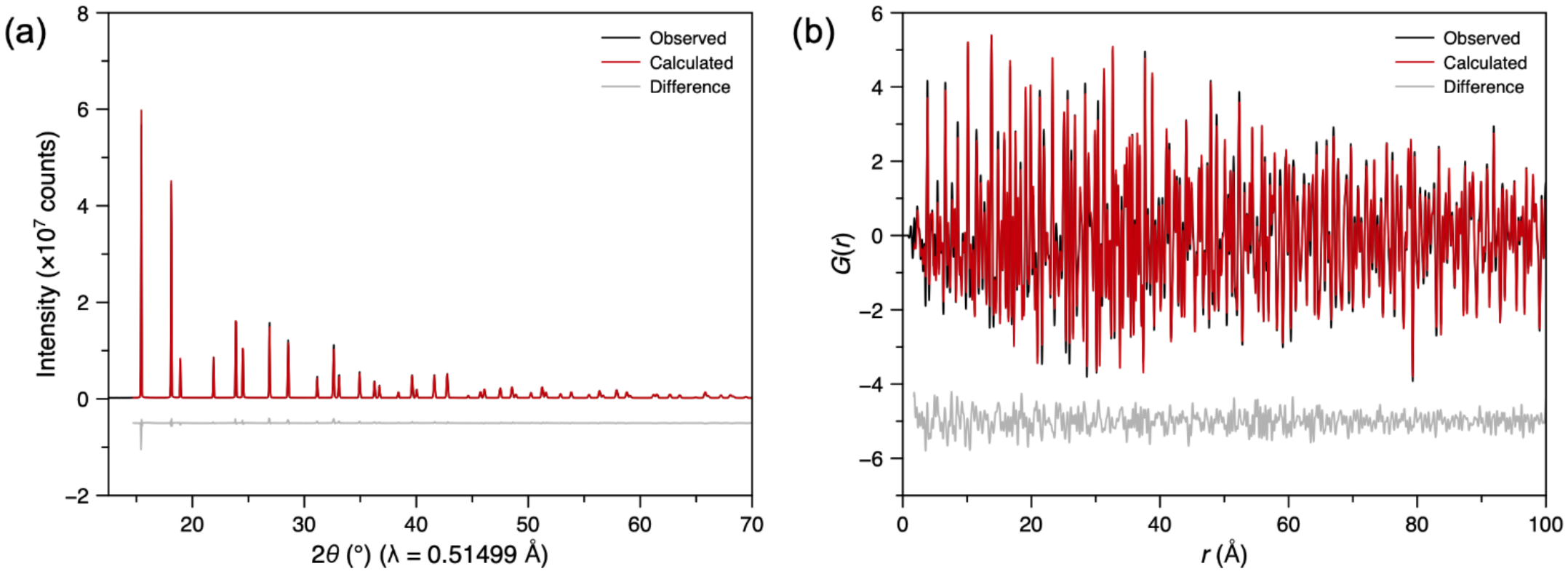

**Figure S7** Refinements against the average data obtained for NIST $CeO_2$ 674. **(a)** Rietveld refinement against the powder diffraction data and **(b)** small-box refinement against the PDF.

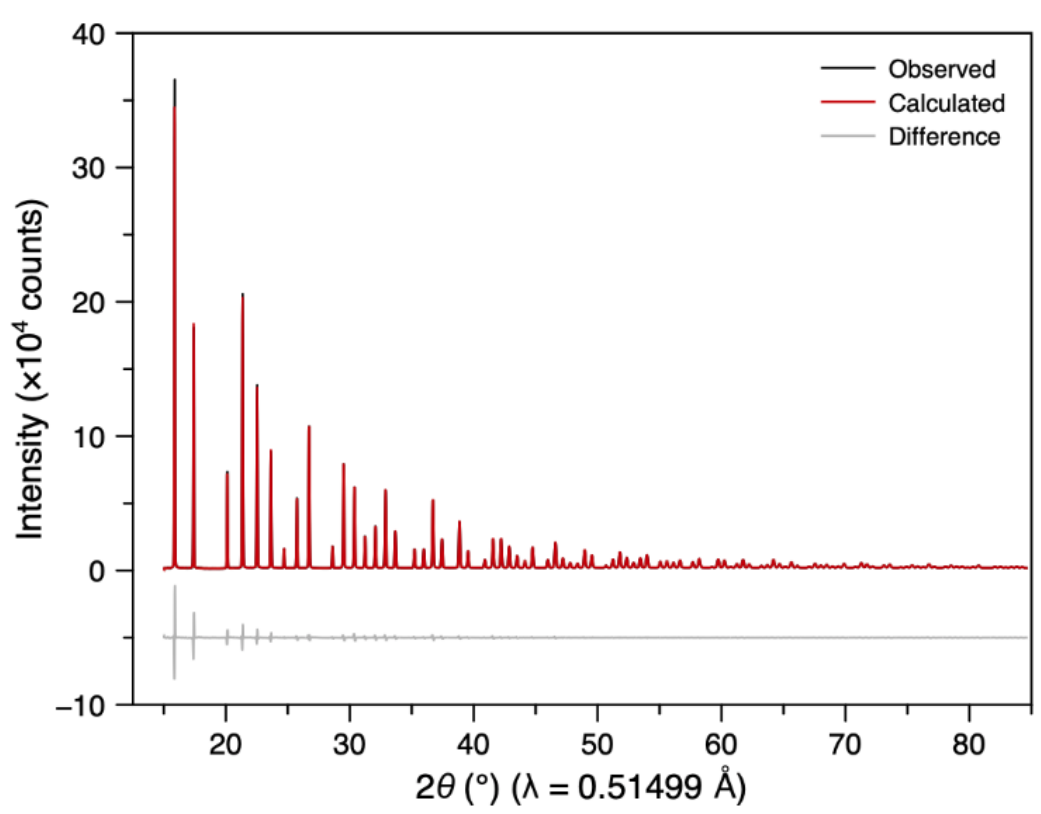

**Figure S8** Rietveld refinement against the average NIST $LaB_6$ 660b diffraction data.

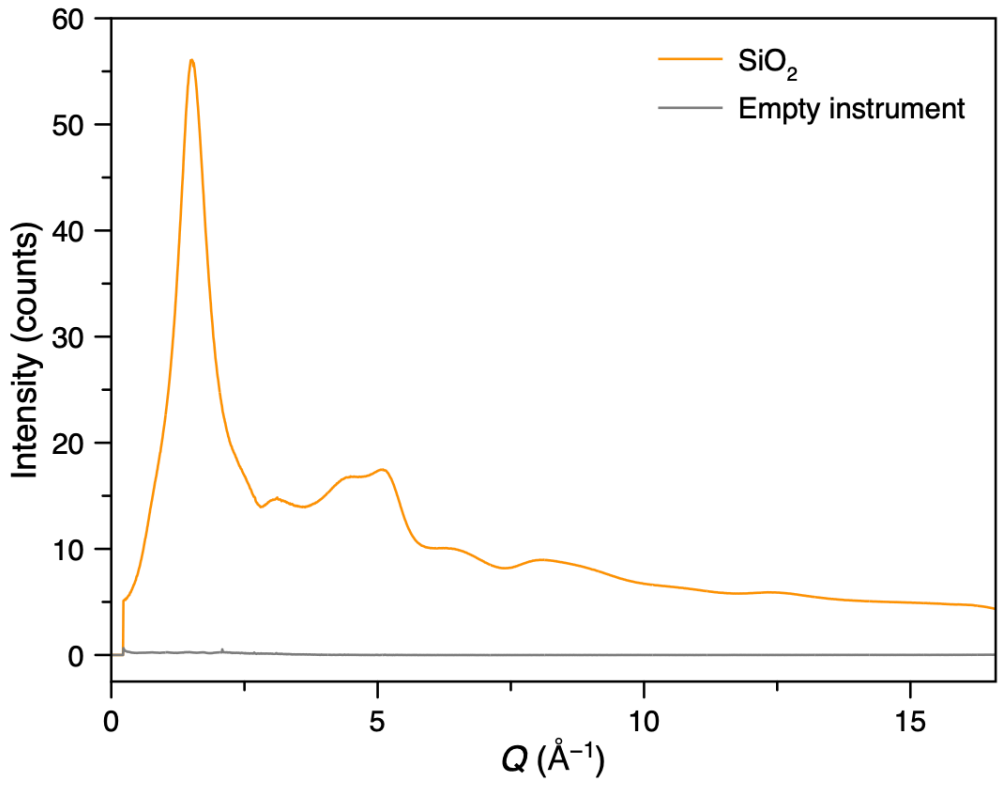

**Figure S9** Comparison between the average $SiO_2$ diffraction data and data collected from the empty instrument.

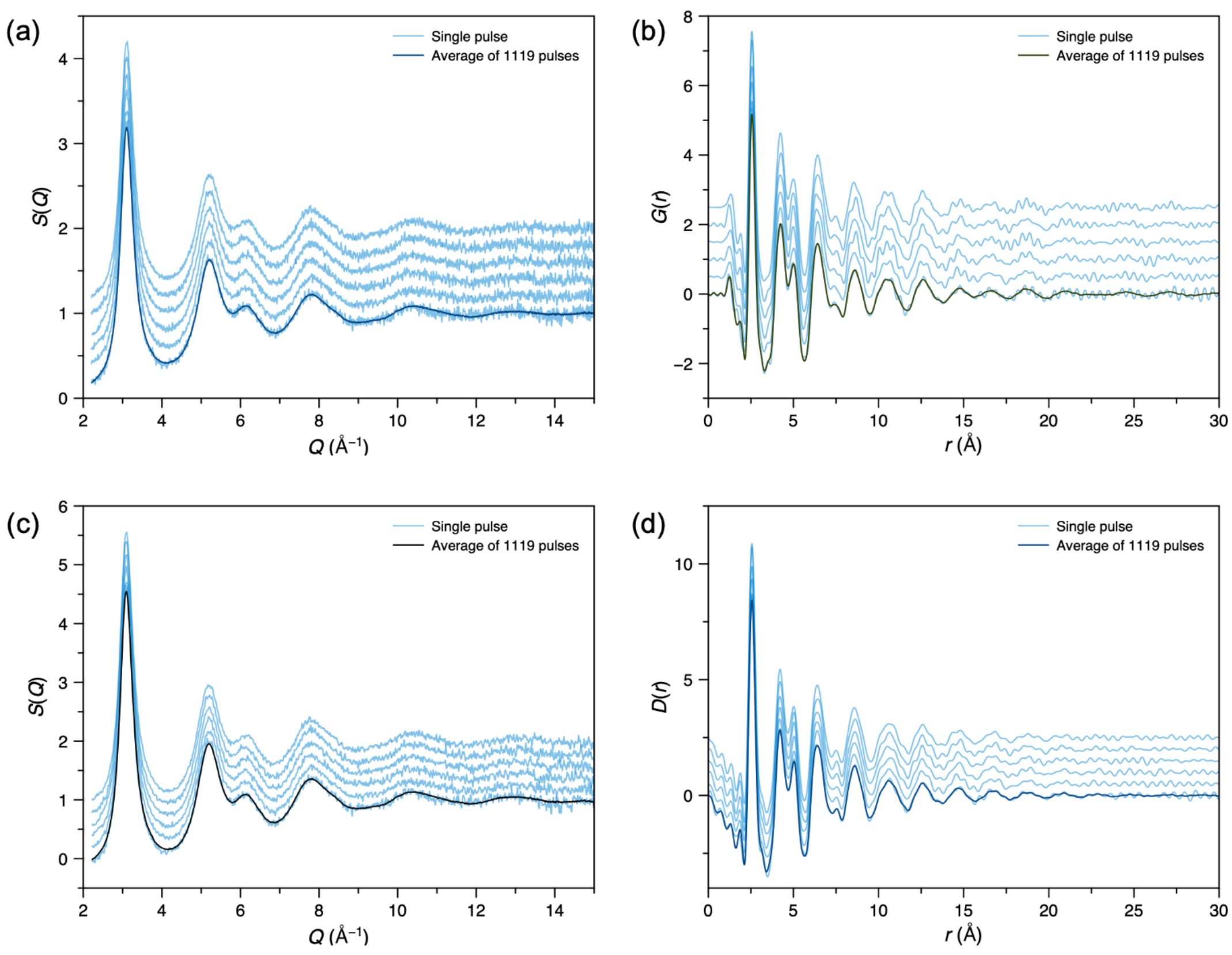

**Figure S10** Representative examples of single pulse data (blue lines) obtained for the metallic glass, compared with the average data (black line). **(a)** and **(b)** show the $S(Q)$ and $G(r)$ obtained from PDFgetX3, respectively. **(c)** and **(d)** show the $S(Q)$ and $D(r)$ obtained from GudrunX, respectively. In (a) – (d), some data sets are offset vertically for clarity.

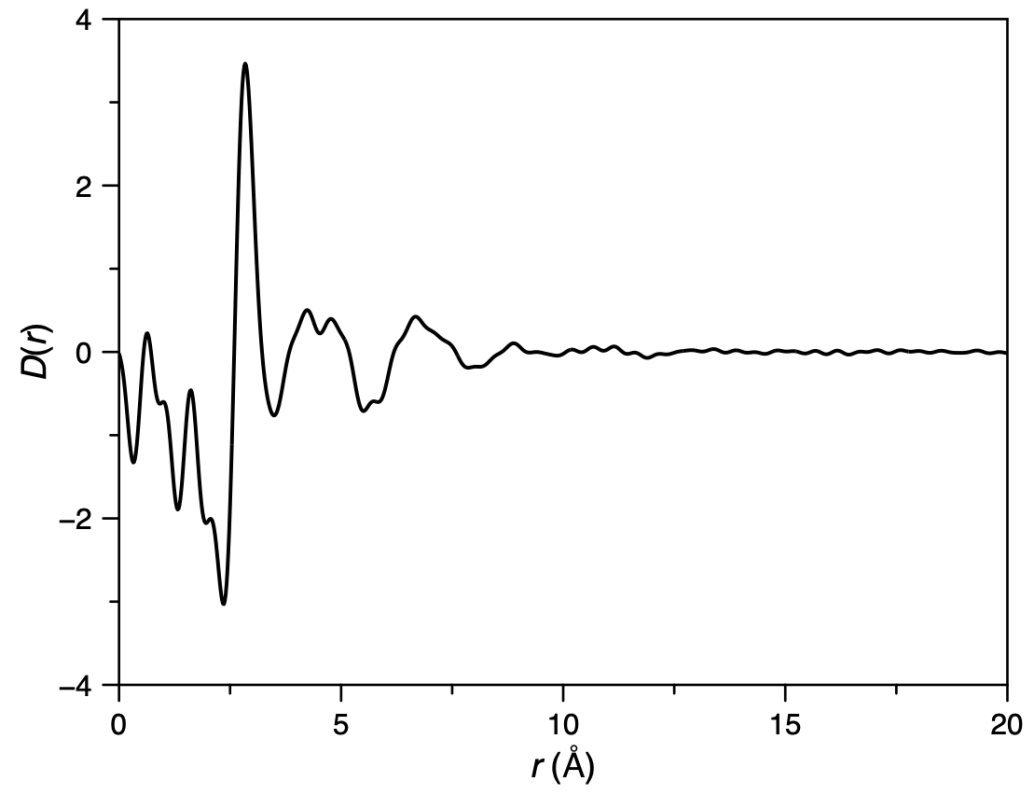

**Figure S11** Average $D(r)$ obtained from water.

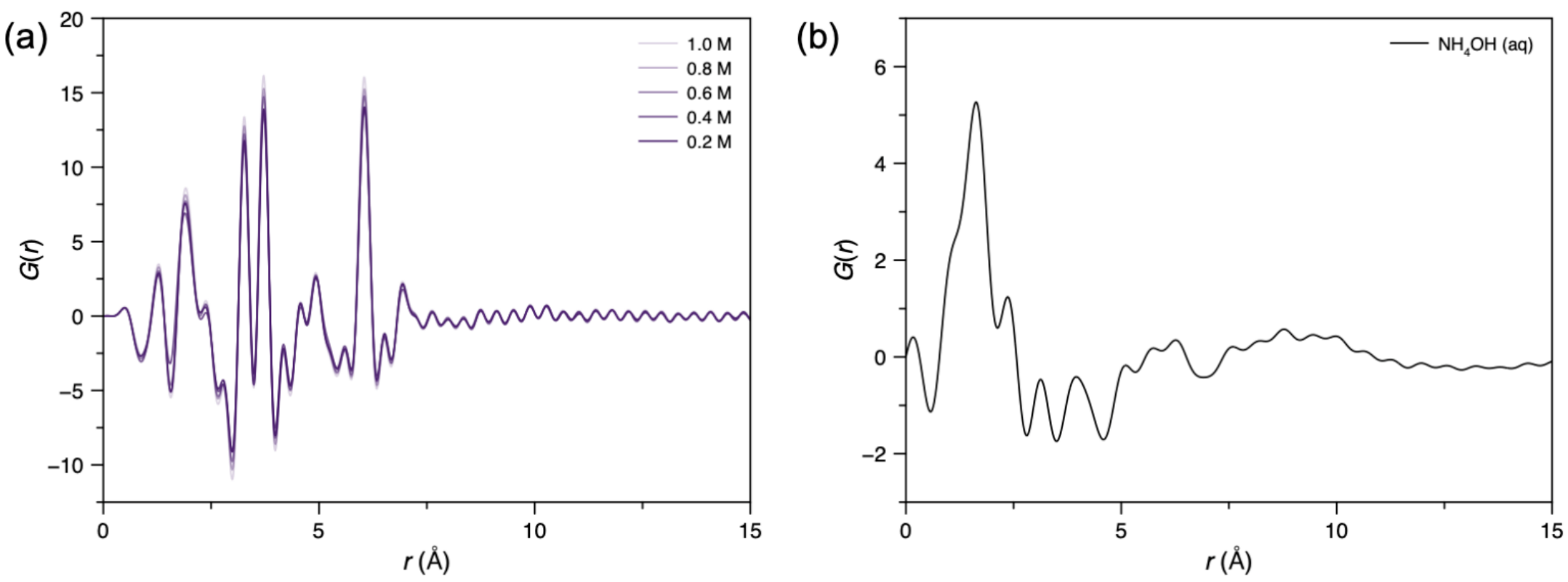

**Figure S12 (a)** Comparison between the difference PDFs of the Keggin cluster at various concentrations and **(b)** reference PDF of $NH_4OH$ (aq), after background subtraction using a capillary loaded with water, measured at a synchrotron to illustrate the solvent-restructuring effects in solution.